\documentclass[journal]{IEEEtran}

\usepackage{epstopdf}
\usepackage[draft]{hyperref} % optional
\usepackage{setspace}
\usepackage{amsmath}
\usepackage{amssymb}
\usepackage{stfloats}
\usepackage{amsthm}
\usepackage{amsfonts}
\usepackage{color}
\usepackage{graphicx}
\usepackage{subfigure}  % for making a row or column of subfigures

\usepackage[flushleft]{threeparttable} 

\usepackage{upgreek}
\usepackage{bm}
\usepackage{nicefrac}

\usepackage{multirow}

\usepackage{cite}
\usepackage{mathtools}
\usepackage{stmaryrd}
\usepackage[mathscr]{euscript}
\usepackage{array}
\usepackage{mathrsfs}

\usepackage{soul}

\newtheorem{theorem}{Theorem}
\newtheorem{lemma}{Lemma}

\begin{document}
\newcommand{\fpm}{f_{\text{pm}}}
\newcommand{\tfpm}{\Tilde{f}_{\text{pm}}}

\author{  Satish~Mulleti, {\it Member, IEEE}, Resham~Yashwanth~Kumar, Laxmeesha~Somappa, {\it Member, IEEE} % <-this % stops a space
	\thanks{\scriptsize The authors are with the Department of Electrical Engineering, Indian Institute of Technology (IIT) Bombay, India. Email: mulleti.satish@gmail.com, laxmeesha@ee.iitb.ac.in}}
%	\thanks{\scriptsize This project has received funding from  }
	
%}
\title{Modulation For Modulo: A Sampling-Efficient High-Dynamic Range ADC}
\markboth{Modulation Instead of Modulo}%
{Shell \MakeLowercase{\textit{et al.}}: Bare Demo of IEEEtran.cls for Journals}
\maketitle

\begin{abstract}
In the realm of high-dynamic range (HDR) analog-to-digital converters (ADCs), the inclusion of a substantial number of quantization bits is highly sought to keep the quantization error low. However, this leads to an elevated bit rate, rendering such ADCs unsuitable for numerous applications. To address this issue, a strategy that combines modulo-folding with a low-DR ADC can be employed to create an efficient HDR-ADC with a relatively modest bit count. Nevertheless, this approach typically necessitates oversampling the input signal, which in turn escalates the overall bit rate. We introduce an alternative approach for achieving HDR-ADC functionality based on phase modulation (PM). In this method, we employ the analog signal committed for sampling to modulate the phase of a sinusoidal carrier signal. This modulation effectively constrains the DR of the resulting signal, allowing for the utilization of a low-DR ADC with fewer bits. We have derived identifiability results, which facilitate the reconstruction of the original analog bandlimited signal from the samples of the PM signal acquired at the Nyquist rate. These results have been extended to accommodate wide classes of signals and are adaptable to non-uniform sampling scenarios. For practical implementation, we propose using discrete phase demodulation algorithms to recover the true samples from the PM samples. Unlike the modulo-folding-based ADC, our proposed approach can operate without the need for oversampling, particularly in noise-free conditions. In the presence of noise, we demonstrate that the proposed PM-based HDR approach exhibits efficiency, characterized by lower reconstruction errors and reduced sampling rates compared to modulo-based HDR-ADCs. We demonstrate the working of the proposed HDR using a rapid hardware prototype and show that signals with DR ten times greater than that of ADC's DR can be reconstructed from the sample measured at the Nyquist rate. This novel framework offers the potential to replace existing high-bit rate HDR-ADCs while maintaining the existing bit rate requirements.
\end{abstract}

\IEEEpeerreviewmaketitle

\section{Introduction}
Sampling involves converting naturally occurring continuous-time signals into sequences. These sequences are then processed by a digital processor, which offers advantages in terms of power efficiency, algorithm flexibility, and resistance to noise. In hardware, analog-to-digital converters (ADCs) are responsible for both the sampling and quantization processes. Three critical parameters for an ADC include the sampling rate, dynamic range (DR), and the number of bits used for quantization. To ensure accurate representation for bandlimited signals, the sampling rate must meet or exceed the Nyquist rate \cite{shannon,nyquist}. It's essential that the signal's DR remains within the ADC's DR to prevent clipping in the samples.

The selection of the number of bits is crucial to minimize quantization errors, with a high-dynamic range (HDR) ADC requiring a greater number of bits for the same level of error. We discuss two commonly encountered scenarios in applications like communication systems, wearable biosensors, wireless sensors, and more, where low bit rates are preferred. In the first scenario, the analog signal being measured typically comprises the desired signal and an unwanted interference, with the interference often having a greater magnitude than the desired signal. To digitally eliminate this HDR interference, the ADC's dynamic range must be sufficiently large to encompass both the desired signal and the interference. With an HDR ADC, a greater number of bits is needed to maintain low quantization errors. In the second scenario, the signal is composed of multiple components, each with varying dynamic ranges. For instance, in radar imaging, the received signal from multiple targets is a combination of delayed and scaled versions of the transmitted signal. The amplitudes of these target components can differ significantly based on the size and distance of each target from the transmitter. In such cases, the ADC's DR should be expansive enough to capture the largest component of the signal, and the quantization resolution should be fine enough to capture the smallest component. Therefore, HDR-ADCs are preferred, but they entail the trade-off of requiring a greater number of bits, which in turn raises the data's bit rate and volume.

A few approaches are available for mitigating quantization errors when dealing with signal components of limited magnitude or when using a low-DR-ADC. For instance, non-uniform quantization techniques, such as companders in combination with uniform quantizers, can enhance the quantization of smaller signal components. However, it's crucial that the ADC's dynamic range remains higher than that of the signal itself. If the objective is to sample a signal with a relatively larger dynamic range using a low-DR-ADC, a straightforward solution involves using an attenuator. The attenuator decreases the signal's dynamic range to match the capabilities of the ADC, which has a lower dynamic range. However, during the attenuation process, signal components with small magnitudes may be attenuated to a point below the quantizer's resolution limit, resulting in the loss of information. A large number of bits is necessary to effectively capture these weak signal components, which results in a high bit rate.

 To utilize a low-DR-ADC without the need for additional bits, a novel approach called the modulo-folding-based HDR-ADC is introduced, as detailed in references \cite{ICCP15_Zhao,unlimited_sampling17,uls_romonov,uls_tsp,uls_hyst,bhandari2021unlimited, eyar_tsp, mulleti2022modulo}. This technique involves an initial step in which analog signals are subjected to folding using a modulo-preprocessing block, ensuring that the dynamic range of the folded signal falls within the capabilities of the ADC. Subsequently, the folded signal is subjected to sampling. In this process, the larger signal components are folded, but the weaker signal components remain unaffected. As a result, unlike the attenuator method, the smaller components can be detected without enhancing the quantizer's resolution. This combination of a modulo-folding circuit followed by a low-DR-ADC allows for the sampling of signals with significantly larger dynamic ranges. Consequently, this combined approach, commonly referred to as the modulo-ADC in subsequent discussions, effectively functions as an HDR-ADC without requiring a substantial increase in the number of bits.

While a modulo-ADC offers high dynamic range (HDR) capabilities with fewer bits, it has drawbacks. First, it samples folded signals, whereas most signal processing algorithms are designed based on true or unfolded samples. In theory, unfolding can be achieved by sampling above the Nyquist rate for bandlimited signals, as discussed in references \cite{uls_identifiability, uls_tsp, eyar_tsp}. However, practical unfolding algorithms require significant oversampling beyond the Nyquist rate to accurately estimate the true samples, as indicated in references \cite{unlimited_sampling17, uls_tsp, uls_romonov, eyar_icassp, eyar_tsp, bhandari2021unlimited, bhandari2022back, mulleti2022modulo, guo_bhandari_sieve_2023, mulleti_mod_fri_icassp23, basheer_lasso_2023}. Generally, the oversampling factor (OF), the ratio of the sampling rate to the Nyquist rate, increases with the noise level and the ratio of the signal's dynamic range to the ADC's dynamic range. With increased oversampling, the bit rate also rises. Multi-channel versions of the modulo-ADCs have been suggested that do not require any oversampling but require more than one modulo-ADC and precise control of their DRs \cite{uls_multichannel, gong_multichannel,guo_bhandari_mc_2023}. The second issue is that the theoretical guarantees and unfolding algorithms are primarily limited to the bandlimited signal model. Expanding these results to signal classes like signals in shift-invariant spaces \cite{unser_50,unser_bspline1,eldar_2015sampling}, finite-rate-of-innovation (FRI) signals \cite{vetterli, bluspmag, fri_strang, eldar_sos, mulleti_kernal}, sparse signals \cite{tao2005uncertainty,donoho_cs,candes_spmag}, multiband signals \cite{lin_ppv_98, herley_wong, venkataramani2000,mishali_2009,mishali_2010,mishali2011xampling}, and others may require substantial effort. Lastly, the practical implementation of modulo-ADCs may be constrained to low frequencies and could demand large power sources \cite{uls_tsp,uls_hyst,bhandari2021unlimited,bhandari2022back,mulleti2023hardware}. This bandwidth limitation is attributed to using a feedback loop in the modulo-folding circuit. To fold high-frequency signals, the feedback circuit must be capable of tracking rapid input signal variations, necessitating fast and precise components.

In this paper, we introduce an innovative approach to HDR-ADC based on phase modulation (PM). The fundamental idea is to phase-modulate the signal to be sampled using a sinusoidal waveform with a specific carrier frequency. The amplitude of this sinusoid is adjusted to fit within the DR of the ADC. Subsequently, we uniformly sample this phase-modulated signal, and discrete phase demodulation (DPD) algorithms are employed to recover the original signal samples. With this short summary of the proposed HDR-ADC, our contributions are summarized as follows. 
\begin{itemize}
    \item We establish theoretical identifiability results for the unique recovery of true samples from PM samples in our proposed framework. These results demonstrate that unique recovery is achievable under specific conditions related to carrier frequency and PM index when the sampling rate exceeds or equals the Nyquist rate for bandlimited signals. Unlike modulo-sampling frameworks, our guarantees are perfect and don't require constant factors.

    \item We extend these identifiability results to a broad range of signal classes, illustrating that the proposed PM approach allows for DR restriction while still maintaining a sampling rate equal to or greater than the minimum required rate for the given signal class. This showcases the adaptability and flexibility of our framework.

    \item We introduce two DPD algorithms: one is based on direct inversion using the $\sin^{-1}$ function, and the other relies on discrete Hilbert transform operations. The $\sin^{-1}$-based algorithm is instantaneous and doesn't depend on the relationship between successive samples, obviating the need for oversampling even in the presence of noise. We also derive a Hilbert-transform-based algorithm to enhance noise robustness, reducing estimation errors with higher sampling rates. Comparative evaluations with unfolding algorithms show that our DPD methods result in lower errors under the same noise level, OF, and ADC's DR.
    
     \item We demonstrate the practical application of our proposed HDR-ADC in the context of electrocardiogram (ECG) signal sampling, which often requires HDR-ADC due to baseline noise. Comparisons with conventional HDR-ADC techniques reveal that our approach leads to lower reconstruction errors at a given quantization level without the need for oversampling. 

    \item  We present a hardware prototype of the modulation-ADC, showcasing its capability to sample bandlimited signals with a DR ten times larger than that of the ADC at the Nyquist rate. The modulation is achieved through a direct digital synthesizer (DDS)-based voltage-controlled oscillator (VCO), which does not require a feedback loop and can operate at high frequencies. Furthermore, we demonstrate that our prototype requires a single low-power source, a more efficient solution compared to modulo-ADCs, which demand multiple and larger power sources.
    
\end{itemize}

The structure of the paper is as follows: Section~\ref{sec:proposed PM} presents our proposed PM-based HDR-ADC, covering theoretical guarantees and algorithms. In Section~\ref{sec:ecg}, we delve into the application of the HDR-ADC for ECG signal sampling. Section~\ref{sec:hw} discusses the hardware prototype and its results, followed by concluding remarks.

\section{Problem Formulation}
\label{sec:problem_formulation}
Consider a finite-energy, real-valued, bandlimited signal $f(t) \in \mathrm{B}_{\omega_m}$ whose maximum amplitude is bounded; that is, there exists a finite $c$ such that $|f(t)|\leq c$. The goal is to sample the signal through an ADC whose dynamic range is $[-\lambda, \lambda]$ for some $\lambda>0$. Typically, it is assumed that $c \leq \lambda$ to avoid signal clipping. However, it's important to acknowledge that in practical scenarios, the condition $c \leq \lambda$ may not always be met. Henceforth, we proceed without the assumption that $c \leq \lambda$.

To address situations where $|f(t)| > \lambda$, the signal $f(t)$ must undergo analog domain preprocessing to ensure that the resulting signal falls within a dynamic range of $[-\lambda, \lambda]$. We represent this preprocessing step using the operator $\mathcal{M}_\lambda: \mathbb{R}\rightarrow \mathbb{R}$, which maps the original signal $f(t)$ to a signal in the range $[-\lambda, \lambda]$. This operator can take on linear forms, like an attenuator, or it can exhibit non-linear characteristics as modulo-folding \cite{unlimited_sampling17} other operators \cite{eyar_tsp}). The preprocessed signal is then uniformly sampled to obtain the values $\mathcal{M}_\lambda(f(nT_s))$, with a sampling rate of $\omega_s = 2\pi/T_s$, which is equal to or greater than the Nyquist rate $\omega_{Nyq} = 2\pi/T_{Nyq} = 2\omega_m$. To be precise, the sampling rate is given as  $\omega_s = \text{OF}\times \omega_{Nyq}$, and $\text{OF}\geq 1$ represents the oversampling factor. Our objective is to reconstruct the original signal $f(t)$ from the samples $\mathcal{M}_\lambda(f(nT_s))$ while maintaining the OF close to unity.

\section{Proposed PM-Based HDR-ADC}
\label{sec:proposed PM}
Consider a bandlimited signal $f(t) \in \mathrm{B}_{\omega_m}$ where $|f(t)|\leq c <\infty$. The objective is to reconstruct the signal from discrete measurements using an ADC with DR $[-\lambda, \lambda]$ where $c$ could be larger than $\lambda$. To this end, instead of sampling $f(t)$ directly, we consider sampling the following PM signal
\begin{align}
	\fpm(t) = \lambda\, \sin(\omega_c t + \mu f(t)),
 \label{eq:fpm}
\end{align}
where $\omega_c$ is the carrier frequency and $\mu>0$ is the PM index. The DR of $\fpm(t)$ is within that of the ADC, that is, $\fpm(t) \in [-\lambda, \lambda]$,  and it carries information of the signal $f(t)$. Hence, $\fpm(t)$ could be used instead of attenuation or modulo folding. If we can reconstruct $f(t)$ from samples of $\fpm(t)$, then the proposed approach solves the problem of sampling HDR signals with low-DR-ADCs. In the next couple of sections, we present identifiability results for uniquely recovering $f(t)$ from its PM samples and then discuss practical algorithms and noise robustness aspects. 

\subsection{Identifiability Results for Bandlimited Signals}
We derive conditions such that $f(t)$ can be perfectly reconstructed from its uniform samples $\fpm(nT_s)$ in the absence of noise. Specifically, we show how $f(nT_s)$ can be computed from $\fpm(nT_s)$. In the following, we refer to this process as discrete phase demodulation or DPD. If we could do so, then $f(t)$ can be perfectly reconstructed from its samples $f(nT_s)$ provided that $T_s \leq T_{\text{Nyq}}$. It is important to note that the presence of a carrier frequency is imperative for conventional analog-domain phase modulation and demodulation used in communication. In comparison, in the context of sampling, as considered in this work, we can get back $f(nT_s)$ from $\fpm(nT_s)$ with and without the carrier frequency. In both scenarios, one can sample at the Nyquist rate to perfectly recover the true samples, as outlined in the following theorem.
\begin{theorem}[Sufficient Conditions for Bandlimited Signals]
  \label{them:sufficient}
    Consider a bounded bandlimited signal $f(t) \in \mathcal{B}_m$ such that $|f(t)|\leq c$. The signal is uniquely recovered from the uniform samples $\fpm(nT_s)$ in the following scenarios:
    \begin{enumerate}
        \item For $\omega_c = 0$, we have unique recovery if $T_s\leq T_{\text{Nyq}}$ and $\mu \leq  \frac{\pi}{2c}$.
        \item For $\omega_c\neq 0$, we have unique recovery if $T_s\leq T_{\text{Nyq}}$, $\mu \leq  \frac{\pi}{2c}$, and $\omega_c = k\, \omega_s = \frac{2\pi}{T_s}$ where $k \in \mathbb{Z}^+$.
    \end{enumerate}
\end{theorem}
\begin{proof}
    To prove Theorem~\ref{them:sufficient}, we first consider the case $\omega_c = 0$. In this case, the samples of the PM signal are given as $\fpm(nT_s) = \lambda \sin(\mu f(nT_s))$. Let us assume that there exists another signal $\bar{f}(t) \in \mathcal{B}_{\omega_m}$ where $|\Bar{f}(t)|\leq c$ such that its corresponding PM samples satisfy the measurement $\fpm(nT_s)$, that is,
\begin{align}
    \fpm(nT_s) = \lambda \sin(\mu f(nT_s)) = \lambda \sin(\mu \bar{f}(nT_s)).
    \label{eq:suff1}
\end{align}
Then we have that
\begin{align}
    \mu\, \bar{f}(nT_s) = \sin^{-1} \left(\sin(\mu f(nT_s)) \right)
\end{align}
To proceed further, we use the following results from trigonometry, which says that 
\begin{align}
    \sin^{-1}(\sin(\theta)) =  
    \begin{cases}
       \theta , & \quad \text{for} \quad |\theta|\leq \pi/2, \\
       (-1)^k \, \theta + k \pi, & \quad \text{otherwise},
     \end{cases}
     \label{eq:triginometric}
\end{align}
where $k \in \mathbb{Z}$ ensures that $ |(-1)^k \, \theta + k \pi| \leq \pi/2$. 

Since $|f(nT_s)|\leq c$, then the choice $\mu \leq \frac{\pi}{2 c}$ ensures that $|\mu f(nT_s)| \leq \pi/2$ and hence by using the identity in \eqref{eq:triginometric} we have that $\bar{f}(nT_s) = f(nT_s)$. As the samples are measured at or above the Nyquist rate, $f(t)$ is uniquely identified from $f(nT_s)$ and hence, from $\fpm(nT_s)$.

The results can be extended to the case $\omega_c \neq 0$, by noting that
\begin{align}
    \fpm(nT_s) =& \lambda \, \sin(\omega_c nT_s)\, \cos(\mu f(nT_s)) \nonumber \\
         &- \lambda \, \cos(\omega_c nT_s)\, \sin(\mu f(nT_s)).
         \label{eq:suff2}
\end{align}
Then for $\omega_s = \omega_c/k$, \eqref{eq:suff2} have same form as samples with $\omega_c = 0$. 
\end{proof}

Theorem~\ref{them:sufficient} states two sufficient conditions uniqueness of DPD. In both cases, with and without a carrier, we note that the minimum sampling rate in both cases could be the Nyquist rate. Moreover, since the DPD approach used in the proof is instantaneous, and the sampling rate requirements are only for perfect reconstruction of $f(t)$ from $f(nT_s)$, we also conclude that the condition $T_s \leq T_{\text{Nyq}}$ is also necessary. Specifically, by using either $\omega_c= 0$, $\mu \leq \frac{\pi}{2 c}$, or, $\omega_c= k \omega_s$, $\mu \leq \frac{\pi}{2 c}$, the DPD by using $\sin^{-1}(\cdot)$ uniquely recover samples $f(nT_s)$ with no role played by the sampling rate. On the contrary, one must sample above the Nyquist rate to unfold the modulo samples.

The results of Theorem~\ref{them:sufficient} are independent of any specific algorithm. An algorithm may be required to operate above the Nyquist rate, especially in the presence of noise, as discussed next.

\begin{figure}[!t]
    \centering
    \subfigure[$\omega_c = 0$]{\includegraphics[width = 3.3 in]{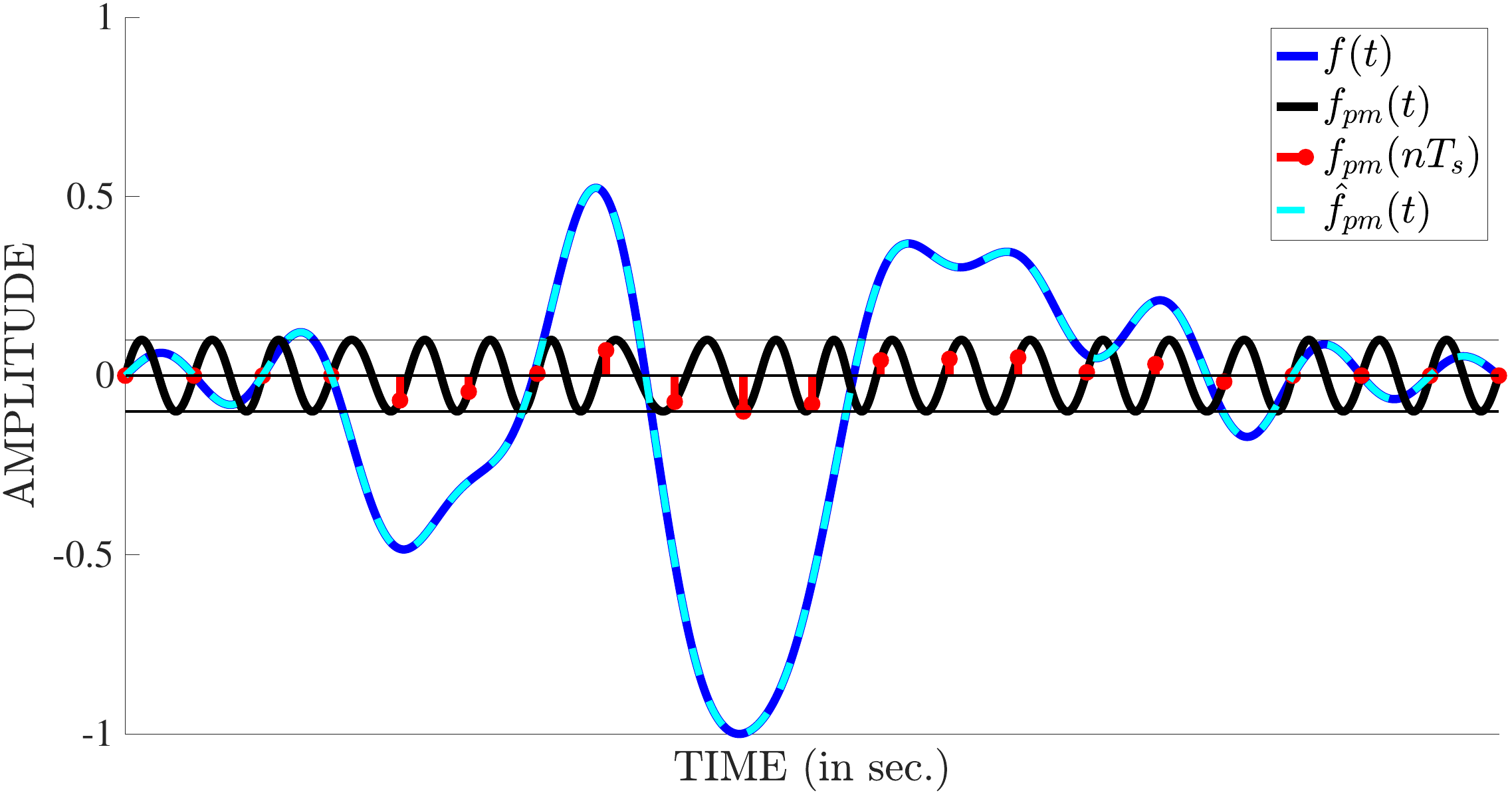}}
    \subfigure[$\omega_c = 2 \omega_s$]{\includegraphics[width = 3.3 in]{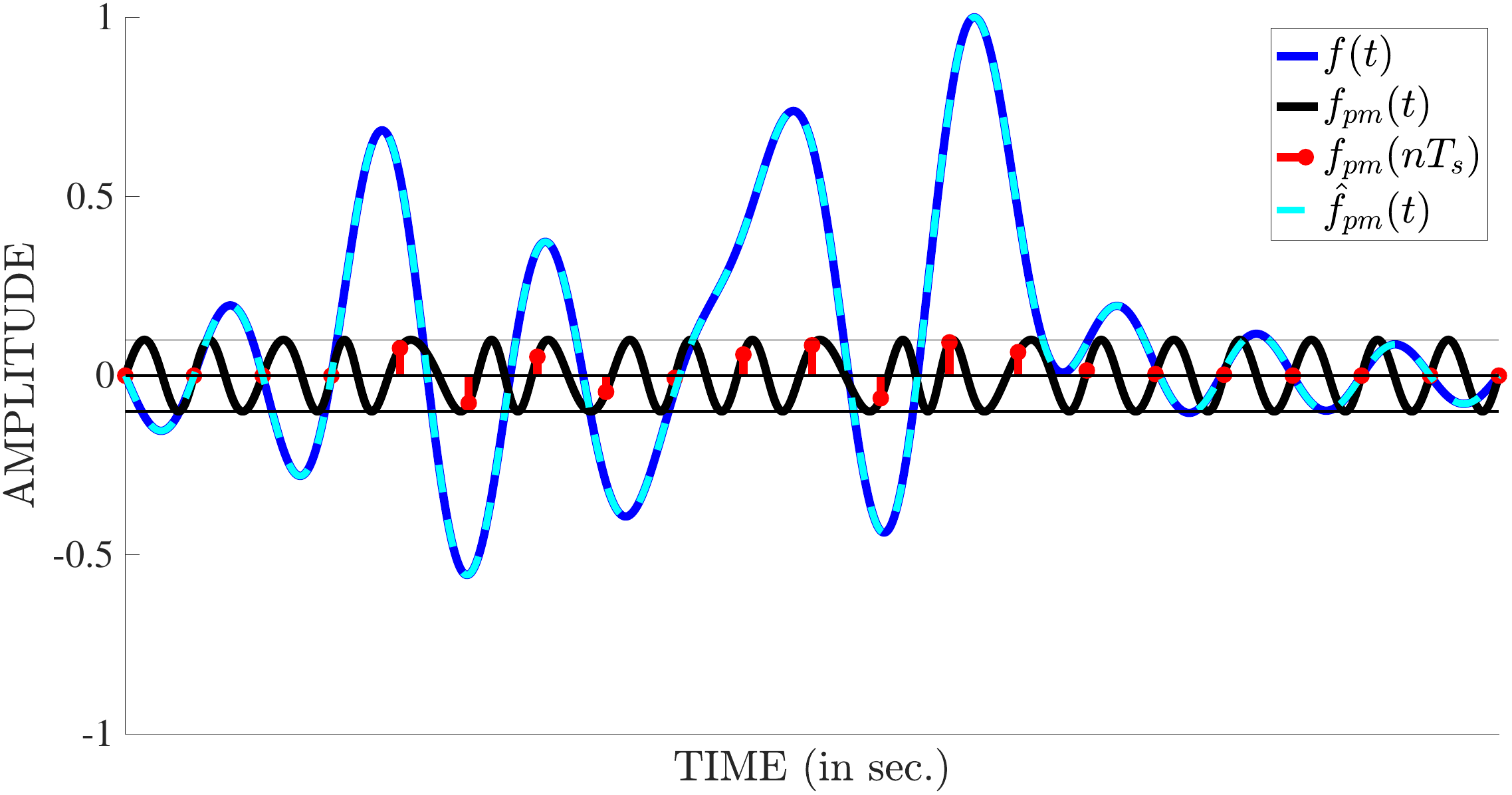}}
    \caption{Illustration of perfect reconstruction of bandlimited signals from the Nyquist samples of their phase modulated signals for (a) with $\omega_c = 0$ and (b) $\omega_c \neq 0$ by using $\sin^{-1}$. The signals are scaled to have a maximum value of one, and the dynamic range of ADC is $\lambda = 0.1$. We are able to perfectly reconstruct signals measured at the Nyquist rate that are 10 times larger than ADC's dynamic range}
    \label{fig:pm_pr}
\end{figure}

\subsection{Identifiability Results for A General Class of Signals and Non-uniform Sampling}
Theorem~\ref{them:sufficient} considers bandlimited signals. Here, we extend the results to general classes of signals. Consider a class of signals $\mathcal{S}_{\boldsymbol{\theta}}$ where $\boldsymbol{\theta}$ denotes a set of parameters of the signals. For example, if $\mathcal{S}_{\boldsymbol{\theta}}$ denotes FRI signals for a given pulse shape $h(t)$, then any $f \in \mathcal{S}_{\boldsymbol{\theta}}$, we can write
\begin{align}
    f(t) = \sum_{\ell=1}^L a_{\ell}\, h(t-\tau_{\ell}),\label{eq:fri}
\end{align}
the number of pulses $L$, the amplitudes $a_{\ell}$s and time delays $\tau_{\ell}$ are parameters of the signals \cite{vetterli, bluspmag, fri_strang, eldar_sos, mulleti_kernal}. A key objective of the FRI sampling problem is to estimate the FRI parameters $\{a_{\ell}, \tau_{\ell}\}_{\ell=1}^L$ from measurements of $f(t)$ by assuming that $h(t)$ and $L$ are known. In the case of shift-invariant space with a basis function $h(t)$, any function $f \in \mathcal{S}_{\boldsymbol{\theta}}$ can be expanded as
\begin{align}
    f(t) = \sum_{k \in \mathbb{Z}} a_k \, h(t- kT), \label{eq:si}
\end{align}
where $T$ is the step size \cite{unser_50,unser_bspline1,eldar_2015sampling}. A similar representation can be extended to other classes of signals, such as multiband signals \cite{lin_ppv_98, herley_wong, venkataramani2000,mishali_2009,mishali_2010,mishali2011xampling}, sparse signals \cite{tao2005uncertainty,donoho_cs,candes_spmag}, and more.

\begin{figure}[!t]
    \centering
    \includegraphics[width =3 in]{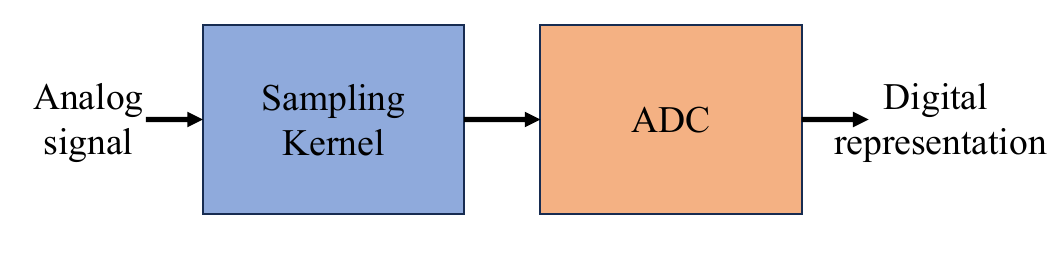}
    \caption{A typical kernel-based sampling framework.}
    \label{fig:general_sampling}
\end{figure}

A typical sampling framework for all the above-mentioned classes of signals is shown in Fig.~\ref{fig:general_sampling}. The analog signal is first passed via an appropriate sampling kernel and then sampled using an ADC, which has sufficient DR and operates above the minimal sampling rate for the signal class. If it is required to use a low-DR ADC, then we can place a PM before the ADC to restrict the DR of the input signal to the ADC. The PM signal can still be sampled at the minimal sampling rate without the PM. The true samples can then be uniquely identifiable by appropriately choosing the carrier frequency and the PM index. The results are summarized in the following theorem.

\begin{theorem}[Sufficient Conditions for General Class of Signals]
  \label{them:sufficient_general}
   Consider a class of signals $\mathcal{S}_{\boldsymbol{\theta}}$ where $\boldsymbol{\theta}$ represents a set of parameters. Let us assume that for any signal $f \in \mathcal{S}_{\boldsymbol{\theta}}$ there exists a sampling kernel $g(t)$ and a minimum sampling interval $T_{\min}$ such that $f(t)$ can be uniquely represented by the uniform samples $(f*g)(nT_{\min})$. Then the signal $f(t)$ is uniquely identifiable from the PM samples denoted as $\lambda \, \sin(\omega_c nT_{s} + \mu (f*g)(nT_s))$ provided that one of the following conditions hold.
   \begin{enumerate}
       \item For $\omega_c = 0$, we have unique recovery if $T_s\leq T_{\min}$ and $\mu \leq  \frac{\pi}{2c}$, where $|(f*g)(t)|\leq c$.
        \item For $\omega_c\neq 0$, we have unique recovery if $T_s\leq T_{\min}$, $\mu \leq  \frac{\pi}{2c}$, and $\omega_c = k\, \omega_{s} = \frac{2\pi}{T_s}$ where $k \in \mathbb{Z}^+$.
   \end{enumerate}
\end{theorem}
The theorem can be proved by following the proof of Theorem~\ref{them:sufficient}.

In this generalized PM-based HDR-ADC, the sampling rate is kept to a minimum while reducing the ADC's DR. The reason for such a straightforward extension of the results from bandlimited signals to general signals is that the DPD is instantaneous and unique for the choices of PM index. Another consequence of this is that the results can be readily extended to a non-uniform sampling framework, as recapitulated below. 

\begin{theorem}[Sufficient Conditions for Nonuniform Sampling]
  \label{them:sufficient_nonuniform}
   Consider a class of signals $\mathcal{S}_{\boldsymbol{\theta}}$ where $\boldsymbol{\theta}$ represents a set of parameters. Let us assume that for any signal $f \in \mathcal{S}_{\boldsymbol{\theta}}$ there exists a sampling kernel $g(t)$ and a set of time instants $\{t_n\}_{n \in \mathbb{Z}}$ such that $f(t)$ is uniquely recovered from the samples $(f*g)(t_n)$. Then the signal $f(t)$ is uniquely identifiable from the PM samples denoted as $\lambda \, \sin(\omega_c t_n + \mu (f*g)(t_n))$ provided that $\omega_c = 0$ and $\mu \leq  \frac{\pi}{2c}$, where $|(f*g)(t)|\leq c$.
\end{theorem}
The proof is similar to that of Theorem~\ref{them:sufficient}.

The results for non-uniform sampling required the carrier frequency to be zero. To derive identifiability conditions for a non-zero carrier frequency, additional constraints on the sampling set $\{t_n\}_{n \in \mathbb{Z}}$ are required. For example, in a sampling-jitter model, the sampling locations are decomposed as $t_n = nT_s + \epsilon_n$, where $T_s$ a fixed sampling interval and $|\epsilon_n| \leq \epsilon < T_s/2$. This sampling pattern captures the non-idealities of an ADC's clock during the uniform sampling. In this case, for $\omega_c= k \omega_s, k \in \mathbb{Z}^+$, we can show that the desired samples can be uniquely identifiable from the PM samples if $|\omega_c \epsilon + \mu c|\leq \pi/2$.

The three theorems presented on identifiability show that different classes of signals can be uniquely recovered from their PM samples without any oversampling. Further, the PM-based approach does not restrict the DR of the signal to be sampled. Specifically, $c$ can take any value, and by adjusting the PM index $\mu$, one can perfectly recover the true samples. 

Next, we discuss practical DPD algorithms.

\subsection{An Algorithm With $\sin^{-1}$}
For practical DPD implementation, we propose using $\sin^{-1}$ operation on the samples $\fpm(nT_s)/\lambda$. This DPD method perfectly recovers $f(nT_s)$ for $\omega_c= 0$, $\mu \leq \frac{\pi}{2 c}$, or, $\omega_c= k \omega_s$, $\mu \leq \frac{\pi}{2 c}$, as discussed in Theorem~\ref{them:sufficient}. In Figs~\ref{fig:pm_pr}(a) and (b), we show the perfect reconstruction of bandlimited signals from the Nyquist samples of PM samples for $\omega_c = 0$ and $\omega_c \neq 0$, respectively. For these simulations, we generated bandlimited signals as
\begin{align}
    f(t) = \sum_{k = -{N_c}}^{N_c} r_k \, \text{sinc} \left( t/T_{\text{Nyq}}-k\right),
    \label{eq:bl_gen}
\end{align}
where $\text{sinc}(t) = \frac{\sin(\pi t)}{\pi t}$. For the simulations, we choose $N_c = 4$, and each coefficient $r_k$ was generated independently from a Gaussian distribution with a mean $0.5$ and variance of four. By setting $T_{\text{Nyq}} = 0.5$ msec., we ensured that the signal was bandlimited to 1 kHz. In the presence of the carrier frequency, we set $\omega_c = 2\omega_s$. For both cases, we selected $\mu = \frac{\pi}{2 c}$. In the two examples, the bandlimited signals are shown in blue, and the corresponding PM signals are in black where $\lambda = 0.1$. The PM signals were sampled at the Nyquist rate of 2k samples/sec. and are depicted in red. For the reconstruction, $f(nT_s)$ were determined from $\fpm(nT_s)$ as $f(nT_s) = \frac{1}{\mu} \sin^{-1} \left(\fpm(nT_s)/\lambda \right)$. In both examples, the reconstructed signals, $\hat{f}(t)$, are shown in cyan, where we observe perfect reconstruction.

The discussed reconstruction algorithm is simple and efficient in terms of sampling rate. However, it may not have favorable robustness for high noise levels. Specifically, when the goal is to estimate the true samples from the noisy samples
\begin{align}
    \tfpm(nT_s) = \fpm(nT_s) + w(nT_s),
    \label{eq:pm_noise}
\end{align}
where $w(nT_s)$ is the noise term, the algorithm estimates $f(nT_s)$ with large errors for high noise levels. To demonstrate the claim, we considered the estimation of noisy samples where $w(nT_s)$ were i.i.d. random variables generated uniformly at random from the interval $[-\sigma, \sigma]$. Since the maximum amplitude of $|\fpm(nT_s)|$ was bounded by $\lambda$, we used the ratio $\sigma/\lambda$ to indicate noise level. Note that $|\tfpm(NT_s)|\leq \lambda+\sigma$, we scaled the PM measurements by $\lambda+\sigma$ before applying $\sin^{-1}$. The estimated samples are given as $\hat{f}(nT_s) = \frac{1}{\mu} \sin^{-1} \left(\tfpm(nT_s)/(\lambda+\sigma) \right)$.

\begin{figure}[!t]
    \centering
    \subfigure[$\sigma/\lambda = 0.1$, NMSE = $-15$ dB]{\includegraphics[width = 3.3 in]{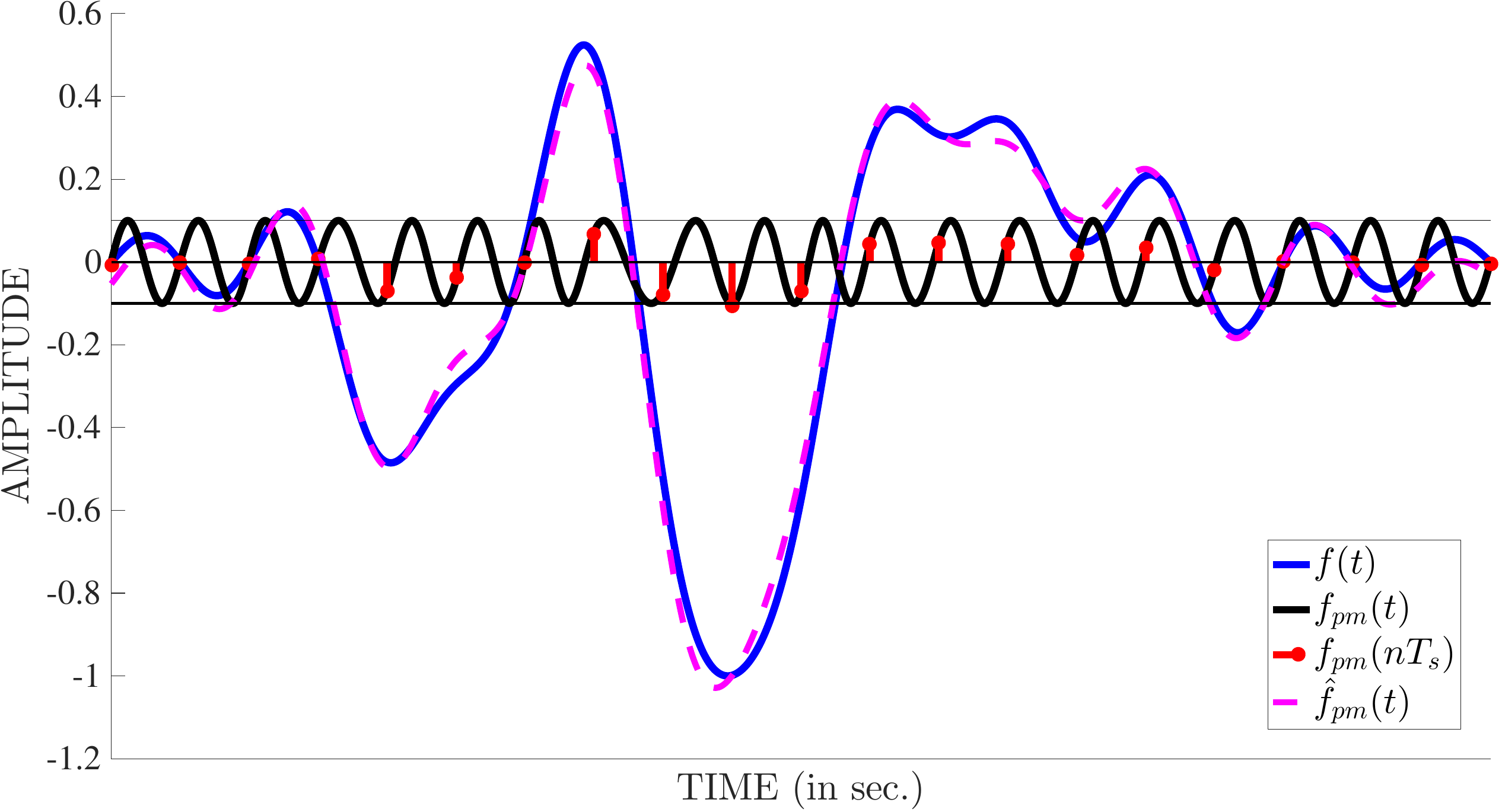}}
    \subfigure[$\sigma/\lambda = 0.3$, NMSE = $-7.2$ dB]{\includegraphics[width = 3.3 in]{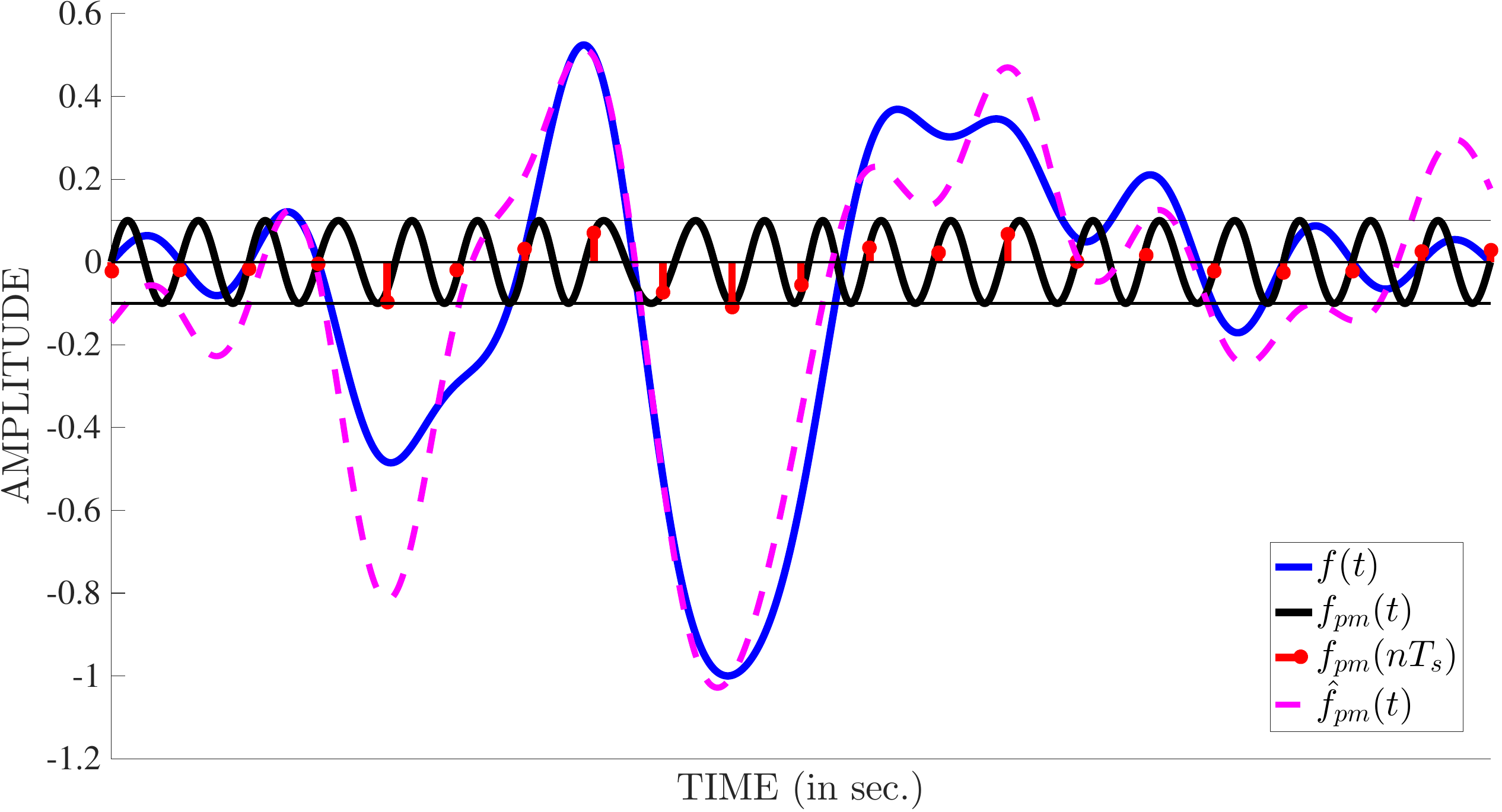}}
    \caption{Reconstruction of signal in Fig.~\ref{fig:pm_pr}(a) by using $\sin^{-1}$ in the presence of bounded noise. Sampling is at the Nyquist rate of the signal.}
    \label{fig:pm_noise}
\end{figure}

In Figs.~\ref{fig:pm_noise}(a) and (b), we show an instance of reconstructions of the bandlimited signal considered in Fig.~\ref{fig:pm_pr}(a) for $\sigma/\lambda = 0.1$ and $\sigma/\lambda = 0.3$, respectively. The samples were measured at the Nyquist rate. We note that the reconstructed signal deviates from the original signal as the noise level increases. To quantify the error, we computed normalized mean-squared error (NMSE) as 
\begin{align}
    \text{NMSE}_f = \frac{\sum_{n} |f(nT_s) - \hat{f}(nT_s)|^2}{\sum_{n}|f(nT_s)|^2}.
    \label{eq:nmse_f}
\end{align}
 An average NMSE calculated from 100 independent noise realizations showed that for $\sigma/\lambda = 0.1$, the algorithm resulted in $-15.2$ dB error, and the error is $-7.2$ dB for $\sigma/\lambda = 0.3$.

 Whether or not the above-mentioned error levels are acceptable for the corresponding noise levels is subjective and depends on the application. However, designing a more robust DPD algorithm is always desirable. In addition, to apply the $\sin^{-1}$ algorithm for $\omega_c \neq 0$, it is required that $\omega_s = \omega_c/k$ where $k \in \mathbb{Z}^+$. The condition could be highly restrictive, and any deviation could lead to large errors. Next, we discuss a robust algorithm for DPD, which is free from the restriction on the sampling rate, as mentioned.

\subsection{A Hilbert Transform-Based Robust Algorithm }
The algorithm we shall discuss is based on applying the discrete Hilbert transform and then estimating the phase of the analytic signal. Specifically, consider samples of the PM signal $f_{\text{pm}}(t)$ with a carrier and without noise given as
\begin{align}
    f_{\text{pm}}(nT_s) = \lambda \, \sin(\omega_c nT_s + \mu f(nT_s)),
    \label{eq:pm_samples}
\end{align}
Let us assume that we have access to the samples of the corresponding quadrature signal
\begin{align}
    f_{\text{pm, quad}}(nT_s) = \lambda \, \cos(\omega_c nT_s + \mu f(nT_s)).
    \label{eq:pm_quad_samples}
\end{align}
Then, we can recover $f(nT_s)$ by constructing the analytic signal and demodulation. Mathematically, by performing the operation
\begin{align}
    \frac{1}{\mu}  \tan^{-1}\left((f_{\text{pm, quad}}(nT_s) + \mathrm{j}f_{\text{pm}}(nT_s))e^{-\mathrm{j}\omega_c nT_s}\right),
    \label{eq:phase_est}
\end{align}
we get back $f(nT_s)$ perfectly provided that $\mu \leq \frac{\pi}{c}$. 

In this perfect reconstruction setup, we do not need the constraint that the carrier frequency should be an integer multiple of the sampling rate. However, in practice, estimation of the quadrature signal $f_{\text{pm, quad}}(nT_s)$ from the samples $f_{\text{pm}}(nT_s)$ is not exact and this results in reconstruction error as discussed next.

For the following discussion, let the operators $\mathscr{H}_c$ and $\mathscr{H}_d$ denote continuous-time and discrete-time Hilbert transforms. Both these transforms have a similar objective, which is to shift the positive frequency content (or for frequencies in the interval $(0, \omega_s/2]$ for discrete signals sampled at a rate $\omega_s$) by $-90^{\circ}$ and the negative frequency content (or for frequencies in the interval $[-\omega_s/2, 0)$ for discrete signals) by $90^{\circ}$. The phase shifts due to the Hilbert transform result in the quadrature components of the signals. For example, we have $\mathscr{H}_c \sin(\omega_0 t) = -\cos(\omega_0 t)$ and $\mathscr{H}_d \cos(\omega_0 nT_s) = \sin(\omega_0 nT_s)$. 

To proceed further, we use the following two results that relate the continuous-time Hilbert transform of a signal to the discrete-time domain counterpart.
\begin{lemma} \label{lem:cht_dht}
    Consider a bandlimited signal $f(t) \in \mathcal{B}_{\omega_m}$. Then we have that 
    \begin{align}
        \mathscr{H}_c f(t)|_{t=nT_s} = \mathscr{H}_d f(nT_s).
    \end{align}
    provided that $T_s \leq \frac{\pi}{\omega_m}$.
\end{lemma}
The results imply that the samples of the CT Hilbert transform are equal to the DT Hilbert transform of the samples, provided that the sampling is performed at or above the Nyquist rate. 

The aforementioned results can be extended to the product of two signals and their Hilbert transform. To elaborate, the Bedrosian theorem \cite{bedrosian1962product} states that for any lowpass signal $f_{L}(t)$ and a highpass signal $f_{H}(t)$ whose spectra do not overlap, we have that
\begin{align}
    \mathscr{H}_c (f_{L}(t)\, f_{H}(t)) = f_{L}(t)\, \mathscr{H}_c f_{H}(t) \label{eq:bedrosian}
\end{align}
If the maximum frequency of $f_{H}(t)$ is $\omega_m$, then following Lemma~\ref{lem:cht_dht}, we have the discrete counterpart of the Bedrosian theorem, which is given as 
\begin{align}
    \mathscr{H}_c (f_{L}(t)\, f_{H}(t))|_{t=nT_s} &= \mathscr{H}_d (f_{L}(nT_s)\, f_{H}(nT_s)), \nonumber \\
    &= f_{L}(nT_s)\,\, \mathscr{H}_d f_{H}(nT_s), \label{eq:d_bedrosian}
\end{align}
provided that $\omega_s \geq 2\omega_m$.

\begin{figure}[!t]
    \centering
    \subfigure[$\sigma/\lambda = 0.1$, NMSE $=-20.6$ dB]{\includegraphics[width = 3.3 in]{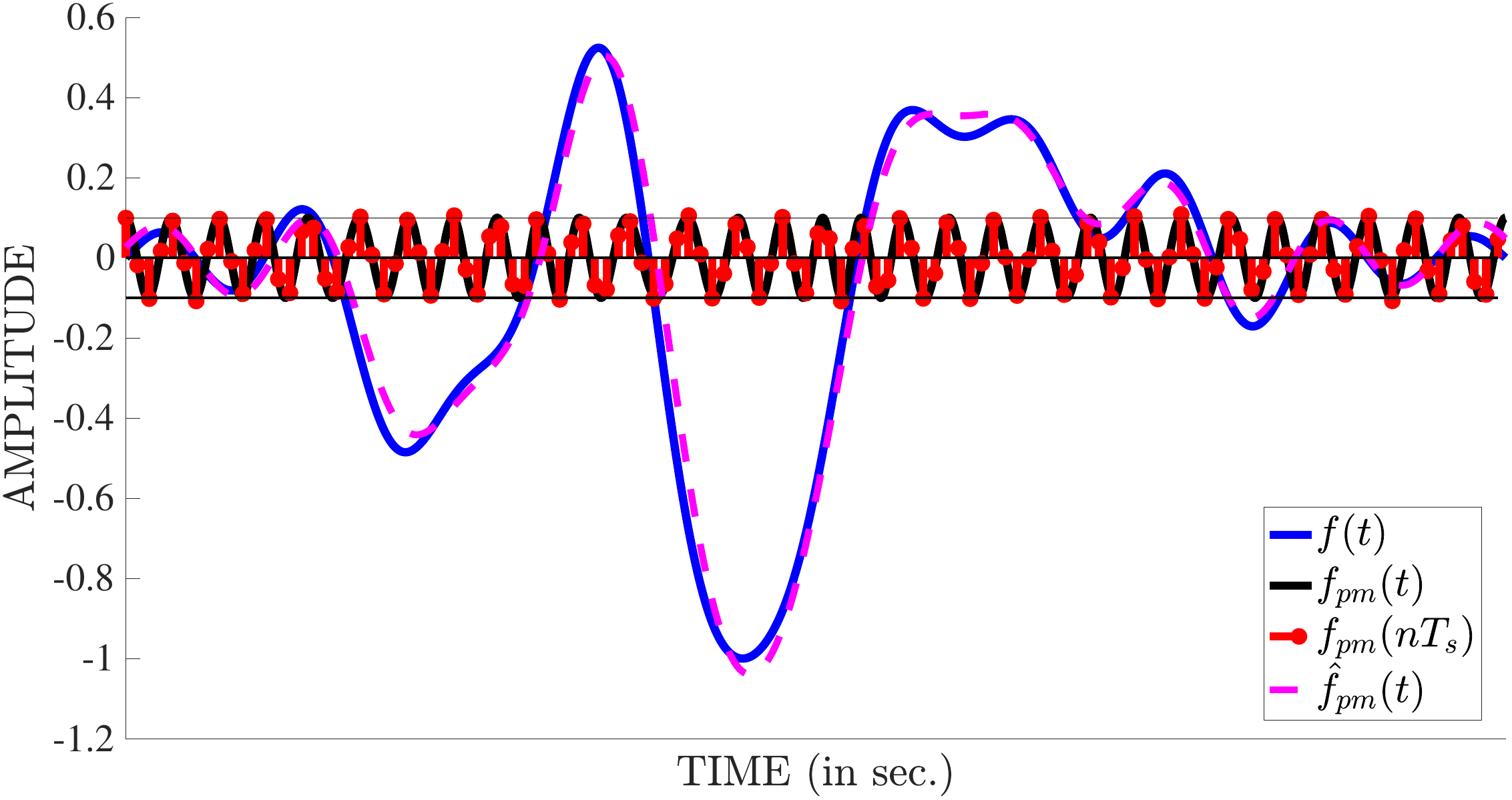}}
    \subfigure[$\sigma/\lambda = 0.3$, NMSE $=-12$ dB]{\includegraphics[width = 3.3 in]{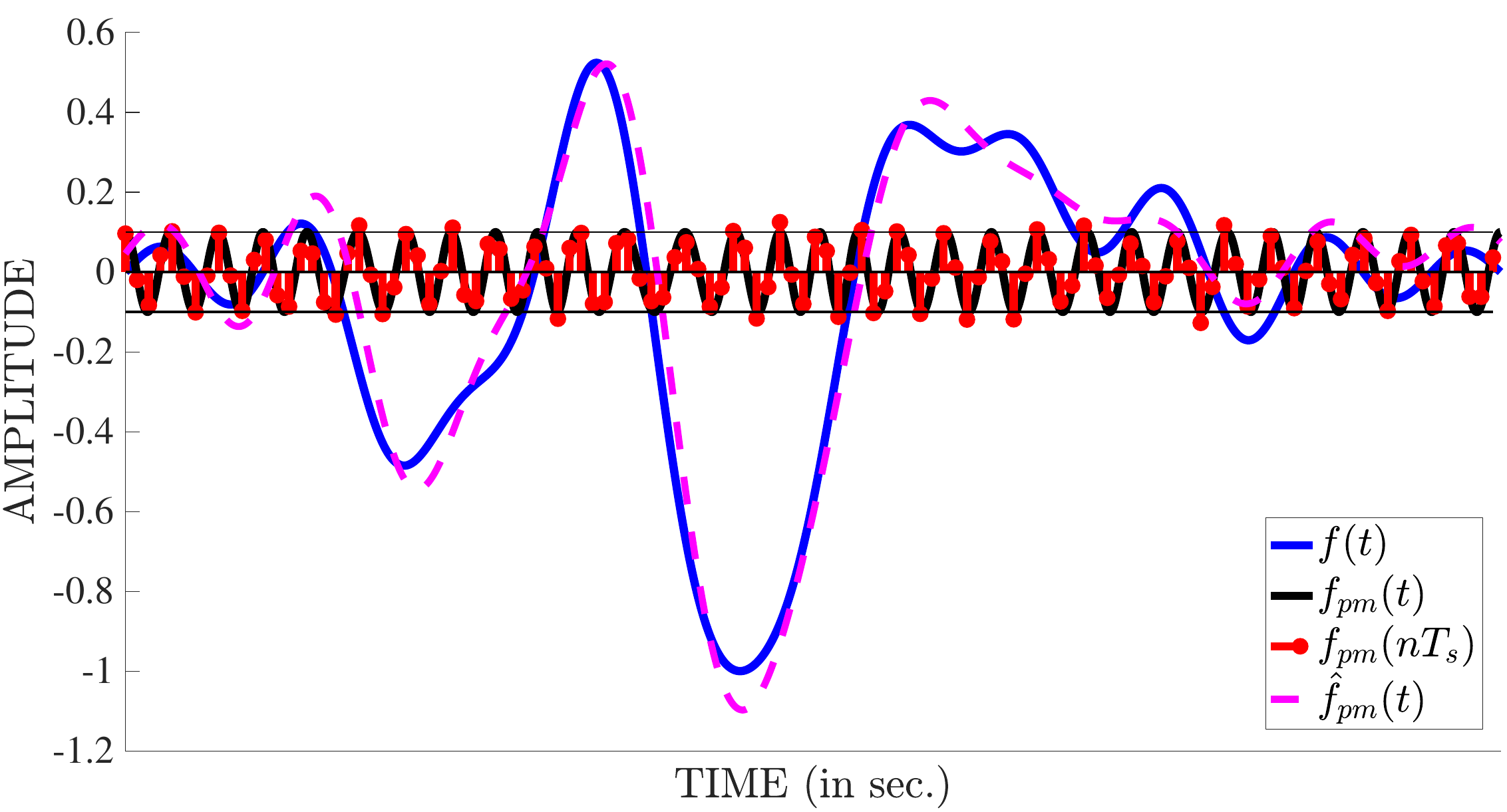}}
    \caption{Reconstruction of signal in Fig.~\ref{fig:pm_pr}(a) by using the discrete-Hilbert transform-based approach in the presence of bounded noise with $\text{OF} = 6$.}
    \label{fig:pm_noise_hilb}
\end{figure}
Given the aforementioned relationship, we can show that $\mathscr{H}_d f_{\text{pm}}(nT_s) \approx f_{\text{pm, quad}}(nT_s)$ provided that the sampling rate is above a threshold. To this end, we first apply the Hilbert transform on $f_{\text{pm}}(t)$ and have the following steps. 
\begin{align}
    &\mathscr{H}_c f_{\text{pm}}(t) = \lambda  \,\,\mathscr{H}_c \sin(\omega_c t + \mu f(t)), \nonumber \\
    & = \lambda \mathscr{H}_c(\sin(\mu f(t))\, \cos(\omega_c t))+ \lambda \mathscr{H}_c(\cos(\mu f(t))\, \sin(\omega_c t)). \nonumber
\end{align}
According to Carson's rule for angle modulation, the functions $\sin(\mu f(t))$ and $\cos(\mu f(t))$ are approximately bandlimited to the frequency interval $[-(\Delta \omega+\omega_m), (\Delta \omega+\omega_m)]$ where $\Delta \omega$ is the maximum frequency deviation given as $\mu \, \underset{t}{\max} \frac{\mathrm{d}f(t)}{\mathrm{d}t}$. Hence, by applying the Bedrosian theorem, we have that $\mathscr{H}_c(\sin(\mu f(t))\, \cos(\omega_c t)) \approx \sin(\mu f(t)) \sin(\omega_c t)$ and $\mathscr{H}_c(\cos(\mu f(t)) \sin(\omega_c t)) \approx -\cos(\mu f(t)) \cos(\omega_c t)$ provided that $\omega_c \geq \Delta \omega + \omega_m$. Hence, we can approximately determine the quadrature signal of the PM signal as
\begin{align}
   \mathscr{H}_c f_{\text{pm}}(t) \approx \lambda \, \cos(\omega_c t + \mu f(t)), 
    \label{eq:ct_quad}
\end{align}
The approximation error depends on the amount of energy of the signals $\sin(\mu f(t))$ and $\cos(\mu f(t))$ outside the frequency interval $[-(\Delta \omega+\omega_m), (\Delta \omega+\omega_m)]$. More specifically, it depends on the spectral overlap between $\sin(\mu f(t))$ and $\cos(\omega_c t)$ or between $\cos(\mu f(t))$ and $\sin(\omega_c t)$. The error can be minimized by choosing $\omega_c$ much larger than  $\Delta \omega+\omega_m$.

By using \eqref{eq:d_bedrosian} and \eqref{eq:ct_quad}, we conclude that
\begin{align}
    \mathscr{H}_d f_{\text{pm}}(nT_s) \approx f_{\text{pm, quad}}(nT_s)
    \label{eq:dt_quad}
\end{align}
provided that the sampling rate is chosen $\omega_s \geq 2\omega_c$. Theoretically, the approximation error in \eqref{eq:dt_quad} will be the same as that in \eqref{eq:ct_quad}. However, in practice, additional errors will result due to the availability of a finite number of samples and due to measurement noise in the PM samples (cf. \eqref{eq:pm_noise}).

The Hilbert-transform-based DPD, suggested here, has seemingly two major drawbacks compared to the $\sin^{-1}$ method discussed in the previous section. First, the sampling rate is higher by an amount $\Delta \omega  = \mu \, \underset{t}{\max} \frac{\mathrm{d}f(t)}{\mathrm{d}t}$. Second, even in the absence of noise, the reconstruction is approximate. Despite these weaknesses, we show that the Hilbert-based method has better noise robustness than the $\sin^{-1}$ method.

To assess the noise robustness of the Hilbert-based reconstruction, we considered the signal in Fig~\ref{fig:pm_pr}(a), and the reconstructed signals are shown in Figs.~\ref{fig:pm_noise_hilb}(a) and (b) for $\sigma/\lambda = 0.1$ and $\sigma/\lambda = 0.3$, respectively. We empirically observed $\Delta f = \mu$ for the signal under consideration where $c = 1$. We set $\mu = 2$,  $\omega_c = (\mu+1)\omega_m$, and $\omega_s = 2\omega_c$ which resulted in an oversampling factor (OF) of $6$ where $\text{OF} = \frac{\omega_s}{2\omega_m}$. We note that the NMSEs for both noise levels were reduced by 5 dB. The sampling rate was set to be minimal. By noting that $f_{\text{pm}}(t)$ was bandlimited to $[-(\Delta \omega+\omega_m), (\Delta \omega+\omega_m)]$, the choice $\omega_s > 2\omega_c$ will lead to separation of spectrum of $f_{\text{pm}}(nT_s)$ and noise over the frequency range $[-\omega_s/2, -(\Delta \omega+\omega_m)] \bigcup [(\Delta \omega+\omega_m), \omega_s/2]$. The separation can be used for denoising by applying a lowpass filter with cutoff frequency $\Delta \omega+\omega_m$. In the following, we use this denoising to compare the proposed HDR-ADC with attenuator and modulo-folding.

\begin{figure}
    \centering
    \includegraphics[width = 3 in]{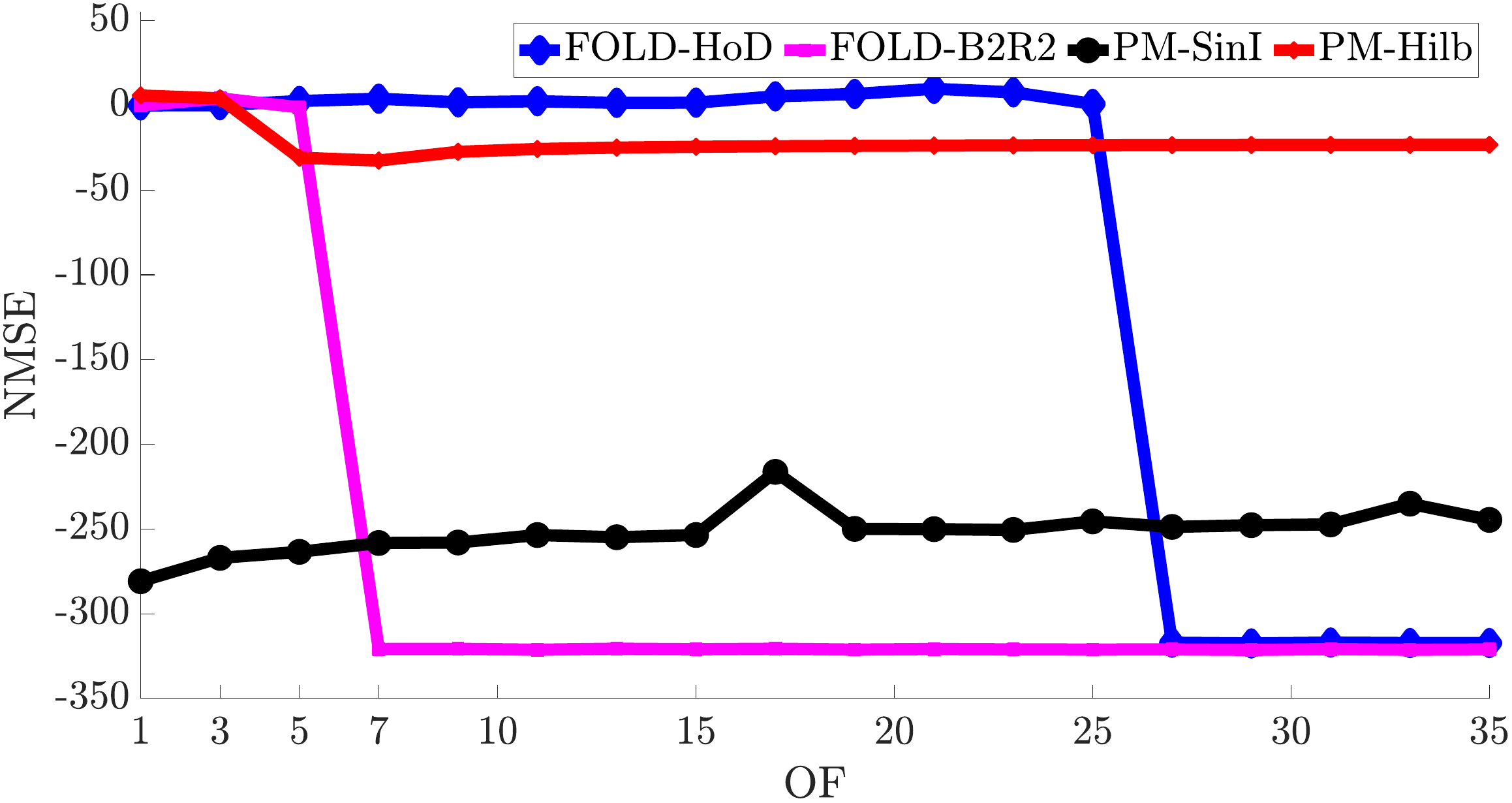}
    \caption{Performance comparison of unfolding and demodulation algorithms for different OFs without noise for $\lambda = 0.05$. $\sin^{-1}$ algorithm does not require any OF. The Hilbert-based approach and unfolding using HoD and B2R2 require $\text{OF}s$ to be greater than 3, 5, and 25, respectively.}
    \label{fig:NMSEvsOF_nonoise}
\end{figure}
\subsection{Performance Comparison of Different HDR approaches}
This section compares the proposed modulation-based HDR with modulo-HDR. To this end, we considered a bandlimited signal shown in Fig.~\ref{fig:pm_pr}(a) where $c = 1$. The signal was used for modulation and modulo-folding for different values of $\lambda$s. The resulting signals were sampled with different OFs. The samples were added with noises of different levels indicated by ratios of $\sigma/\lambda$s. The noises in the simulations were sampled from zero-mean Gaussian distribution with variance $\sigma^2$. Unlike the scenarios depicted in Figs.\ref{fig:pm_noise} and \ref{fig:pm_noise_hilb}, which involved bounded and uniformly distributed noise, our approach here incorporates unbounded Gaussian noise, enhancing the analysis' generality.

\begin{figure*}[!t]
\begin{center}
\begin{tabular}{cccc}
\subfigure[$\lambda = 0.05$, $\sigma = 0.05\,\lambda$]{\includegraphics[width=1.7in]{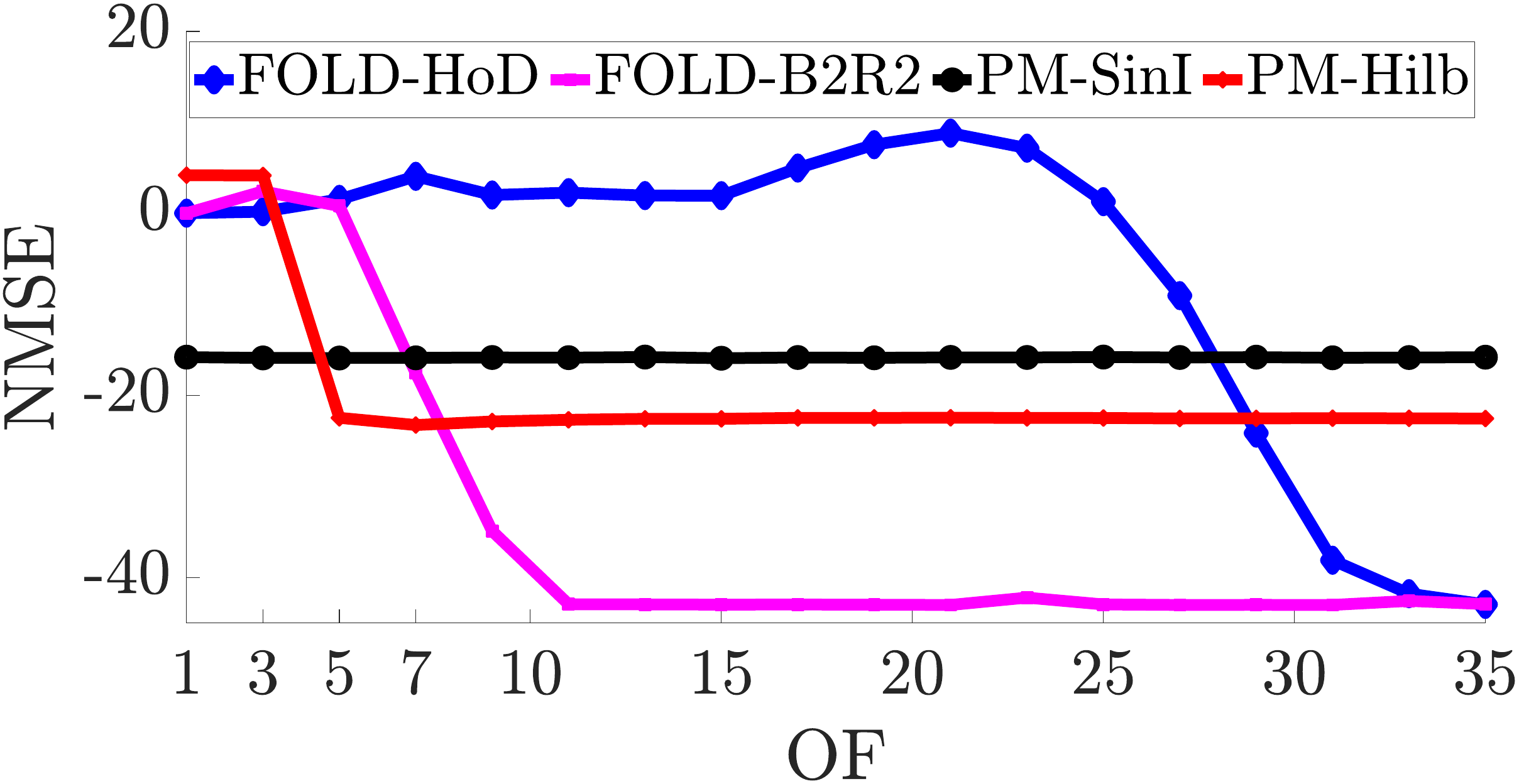}\label{fig:sigma0}} 
\subfigure[$\lambda = 0.05$, $\sigma = 0.1\,\lambda$]{\includegraphics[width=1.7in]{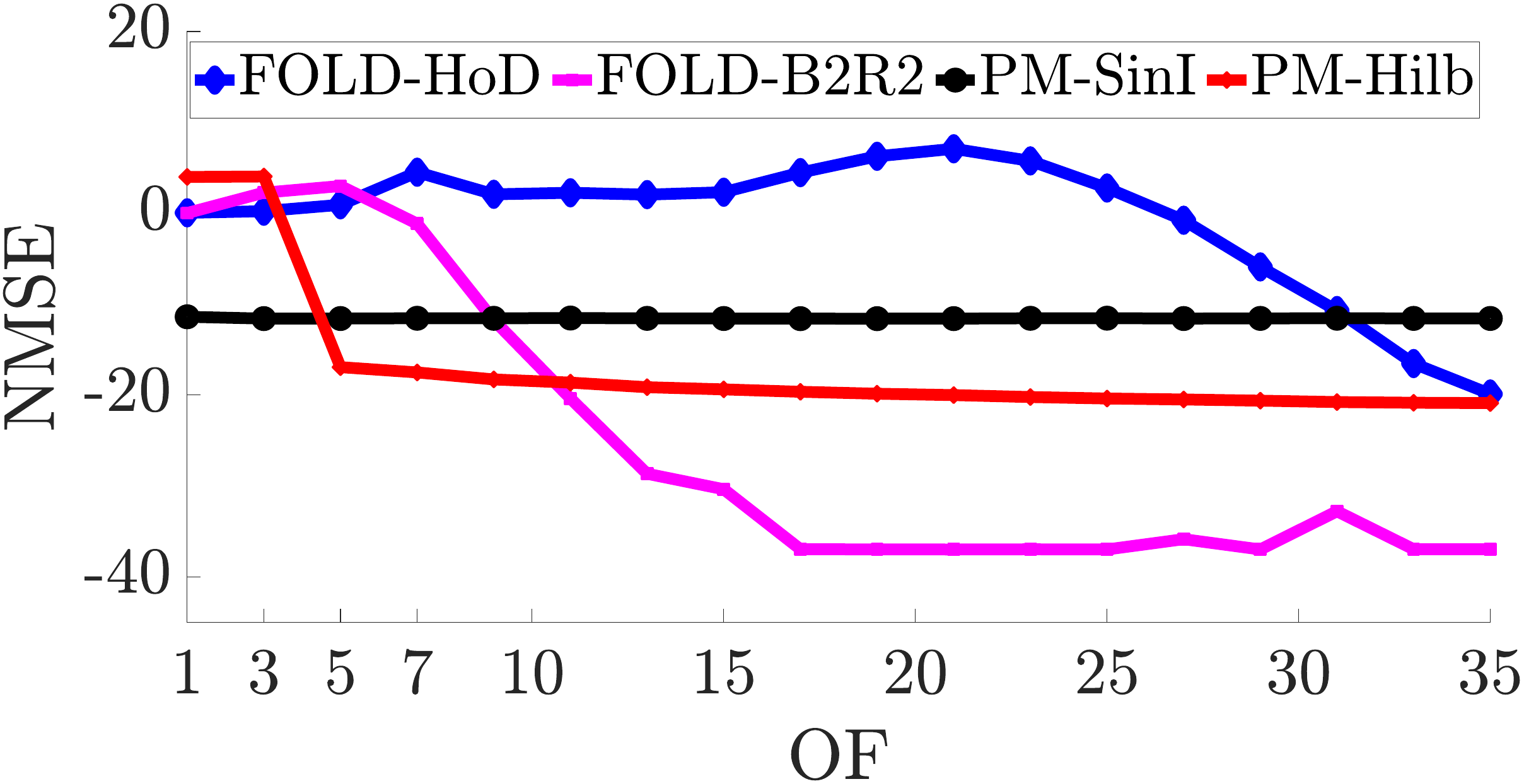}\label{fig:sigma01}} 
\subfigure[$\lambda = 0.05$, $\sigma = 0.2\lambda$]{\includegraphics[width=1.7in]{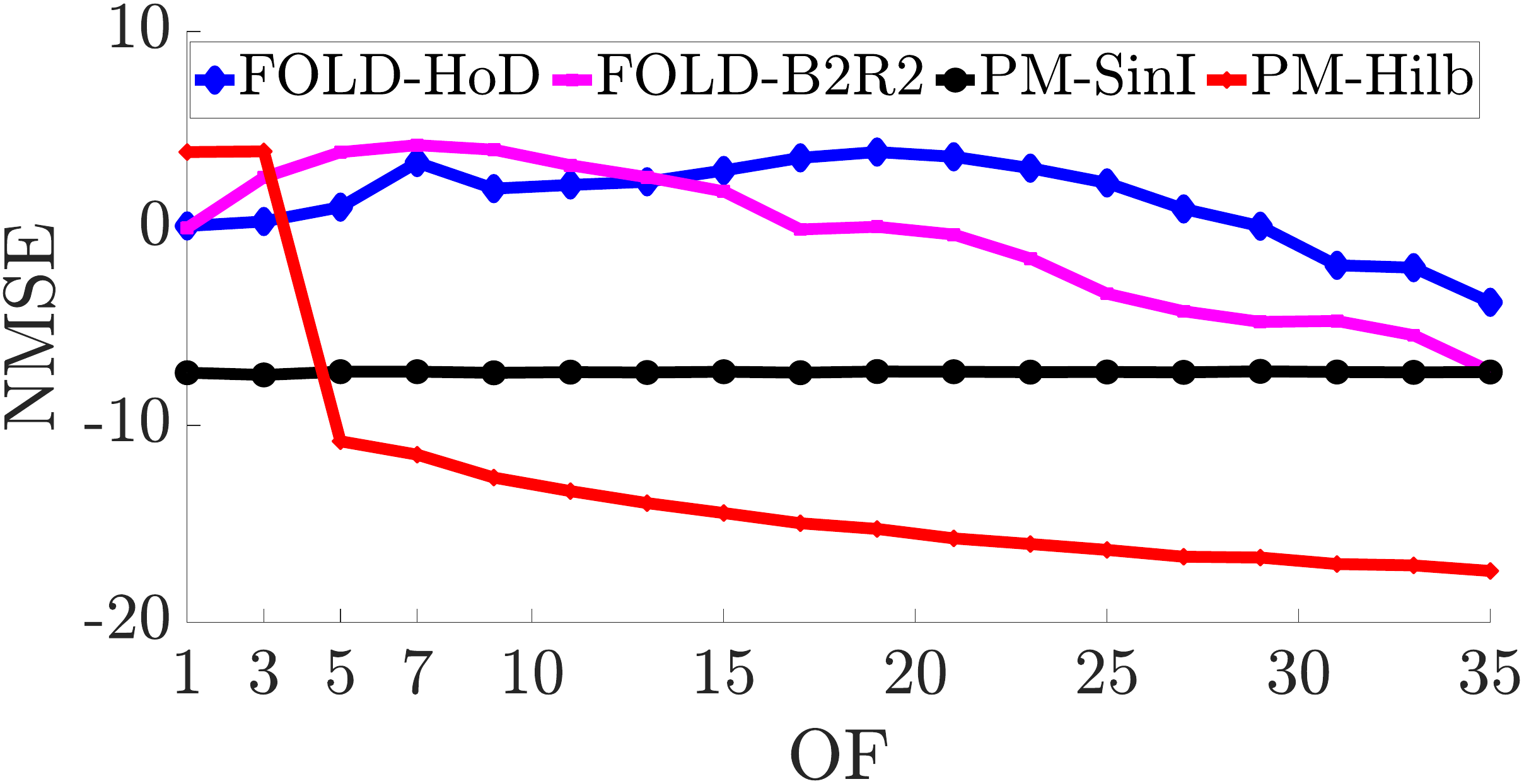}\label{fig:sigma02}}
\subfigure[$\lambda = 0.05$, $\sigma = 0.4\lambda$]{\includegraphics[width=1.7in]{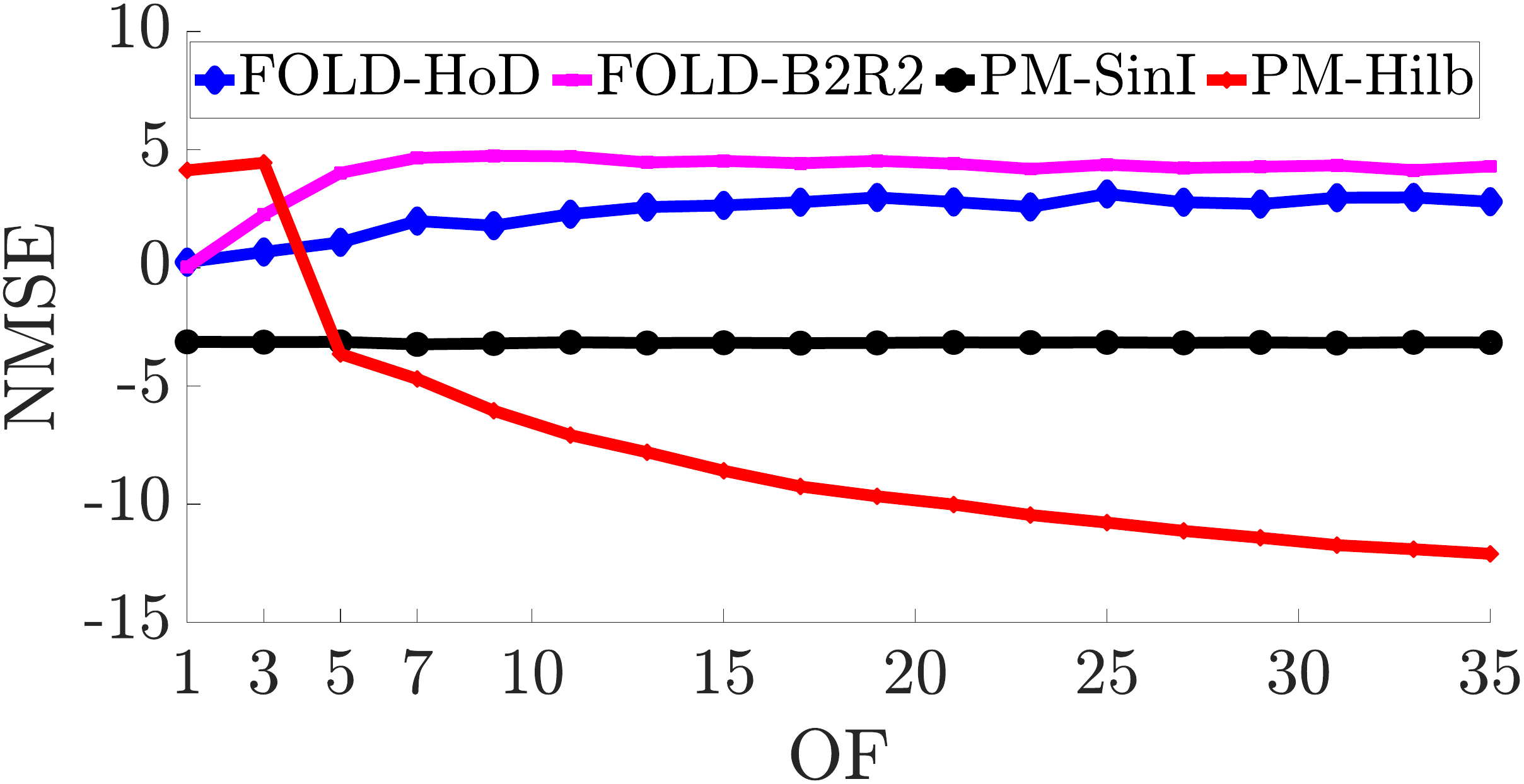}\label{fig:sigma04}}\\
\subfigure[$\lambda = 0.1$, $\sigma = 0.05\, \lambda$]{\includegraphics[width=1.7in]{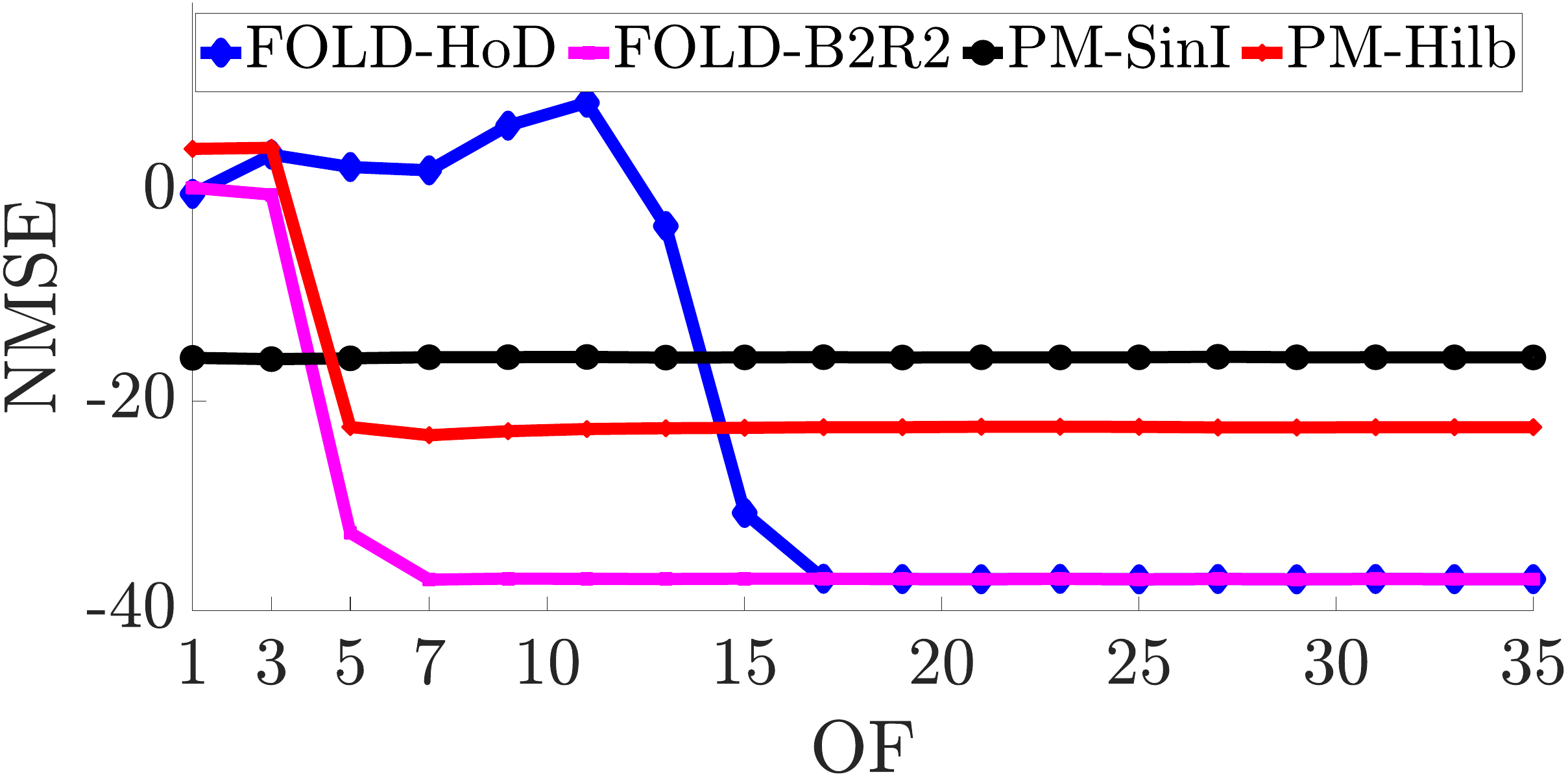}\label{fig:sigma0}} 
\subfigure[$\lambda = 0.1$, $\sigma = 0.1\,\lambda$]{\includegraphics[width=1.7in]{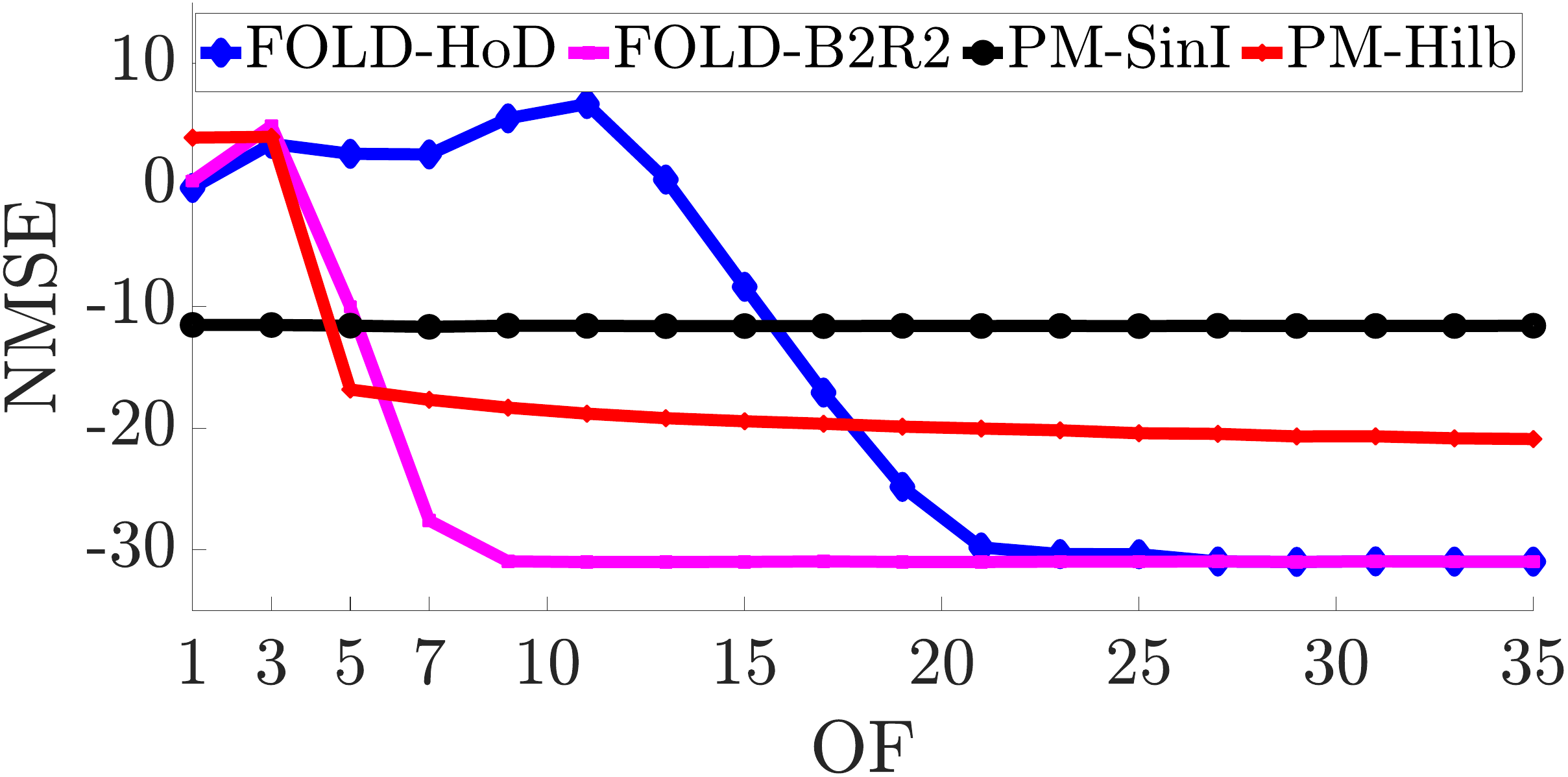}\label{fig:sigma01}} 
\subfigure[$\lambda = 0.1$, $\sigma = 0.2\lambda$]{\includegraphics[width=1.7in]{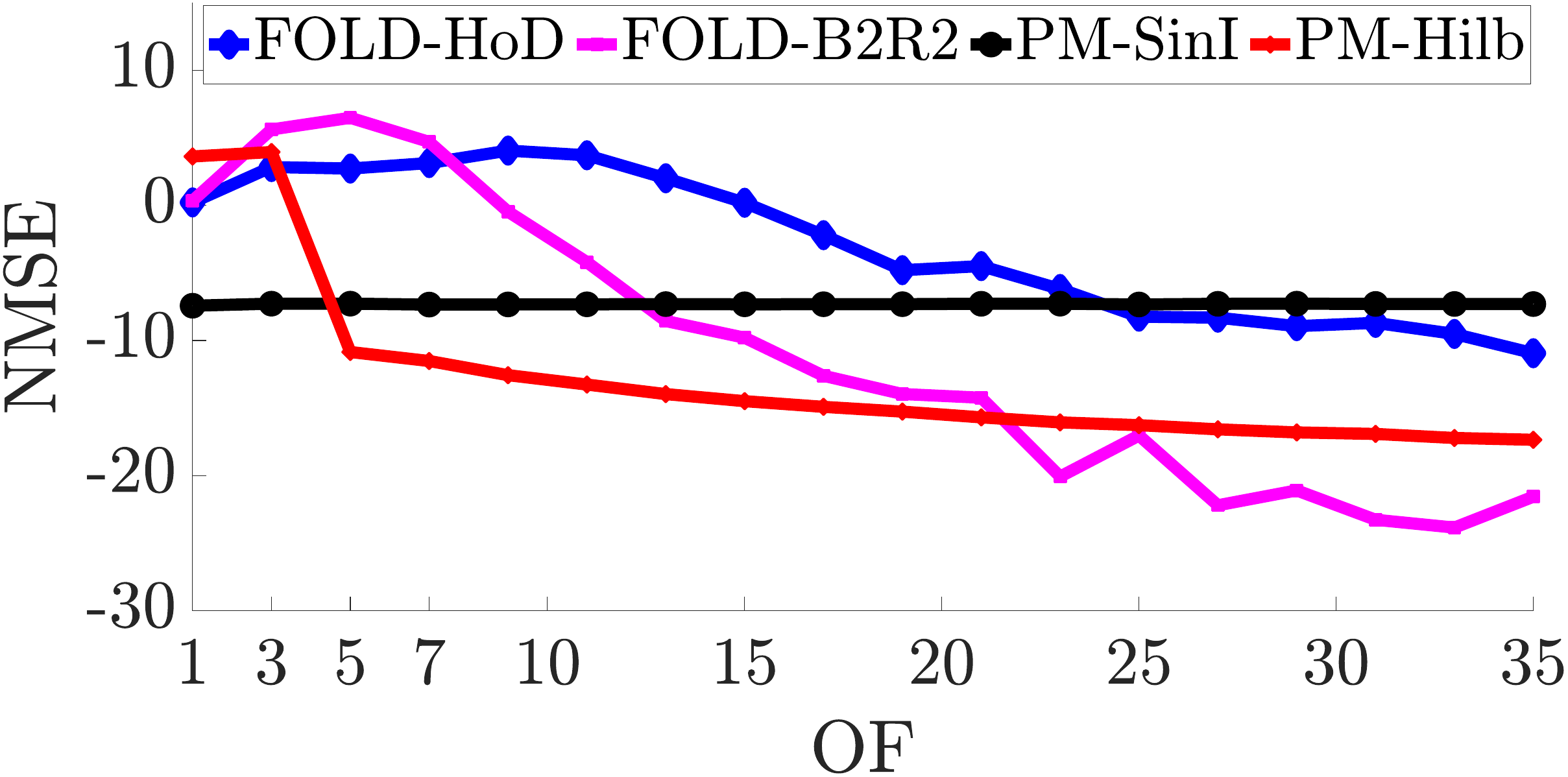}\label{fig:sigma02}}
\subfigure[$\lambda = 0.1$, $\sigma = 0.4\lambda$]{\includegraphics[width=1.7in]{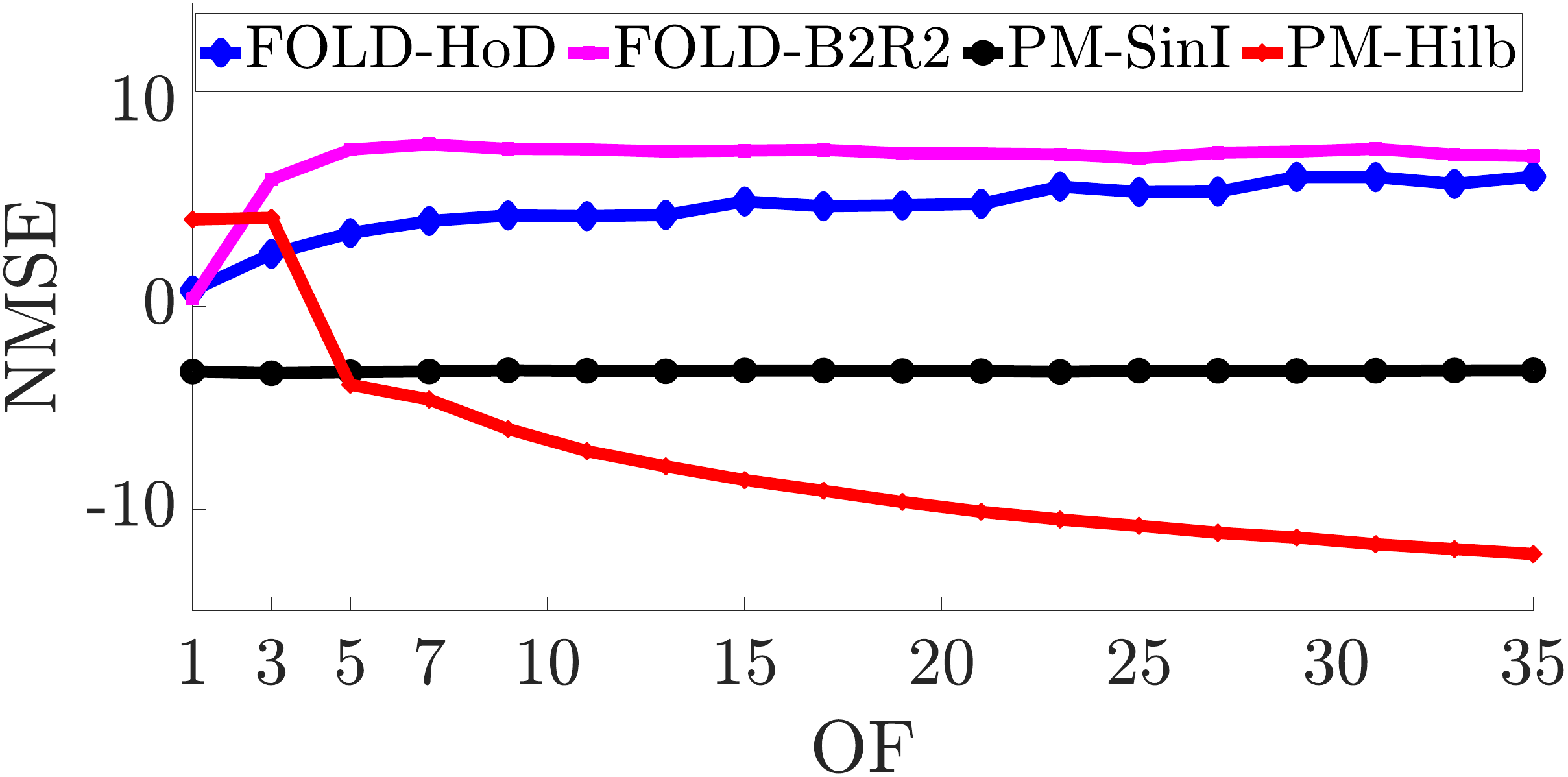}\label{fig:sigma04}}\\
\subfigure[$\lambda = 0.3$, $\sigma = 0.05\, \lambda$]{\includegraphics[width=1.7in]{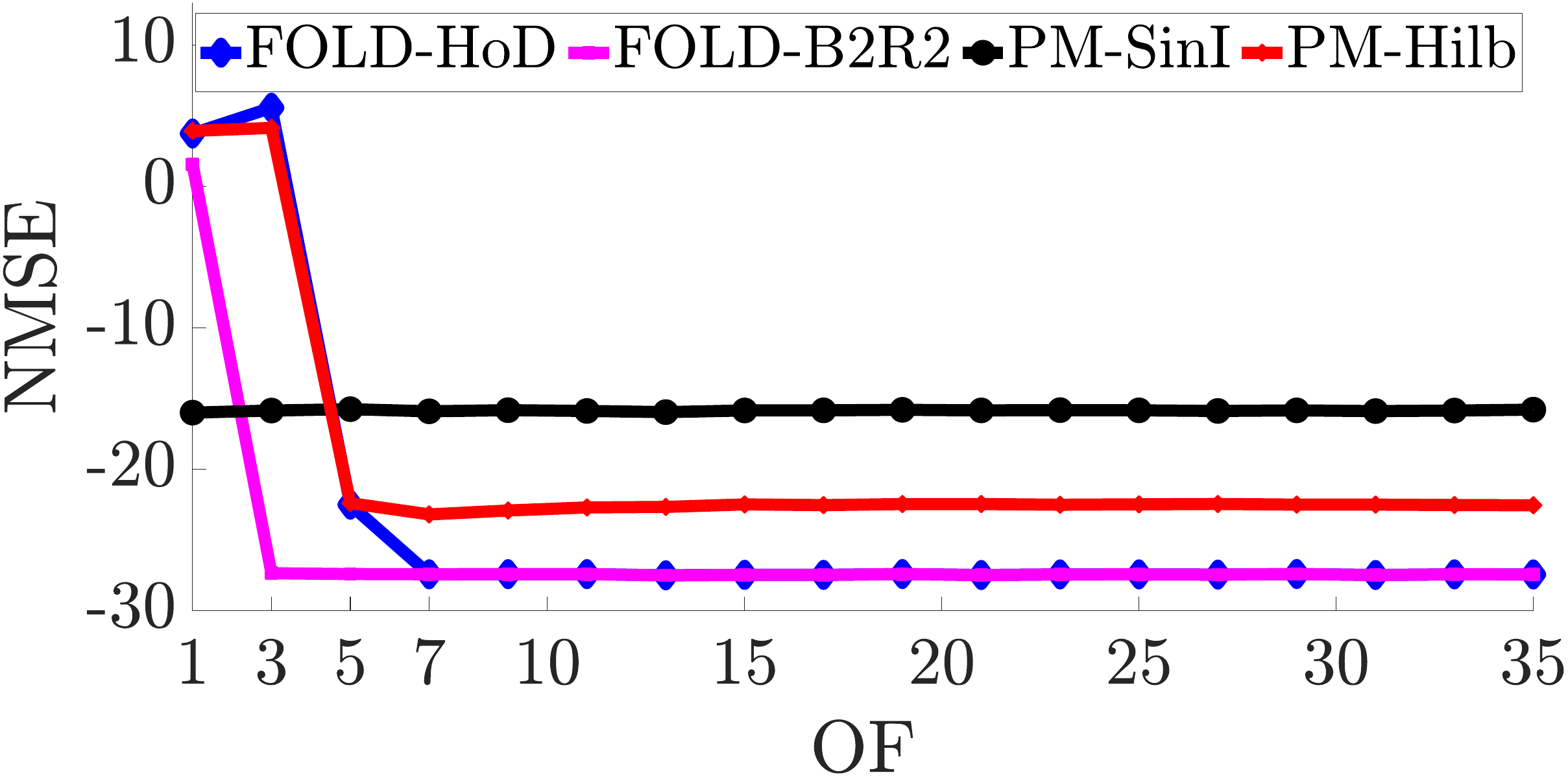}\label{fig:sigma0}} 
\subfigure[$\lambda = 0.3$, $\sigma = 0.1\,\lambda$]{\includegraphics[width=1.7in]{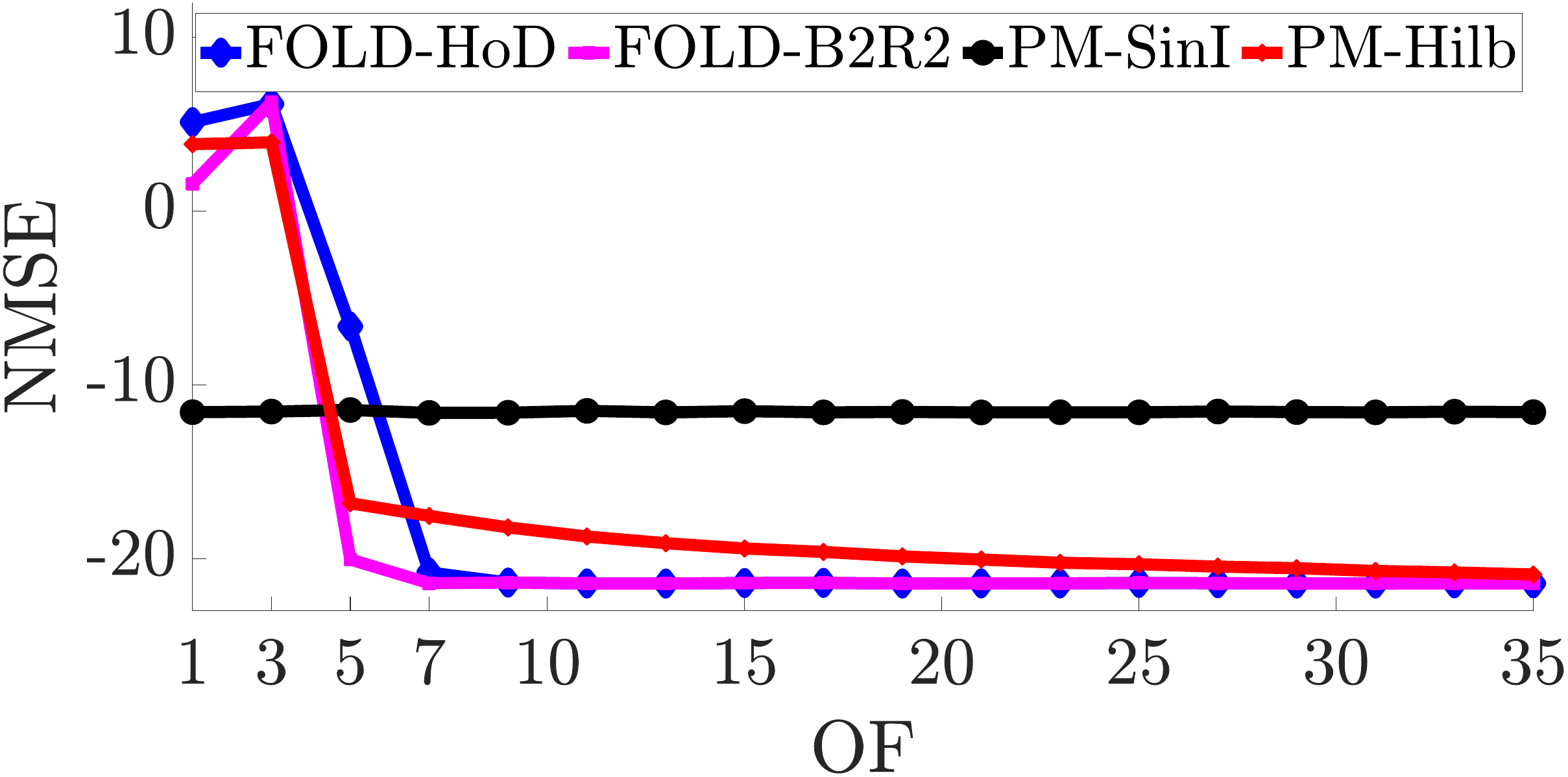}\label{fig:sigma01}} 
\subfigure[$\lambda = 0.3$, $\sigma = 0.2\lambda$]{\includegraphics[width=1.7in]{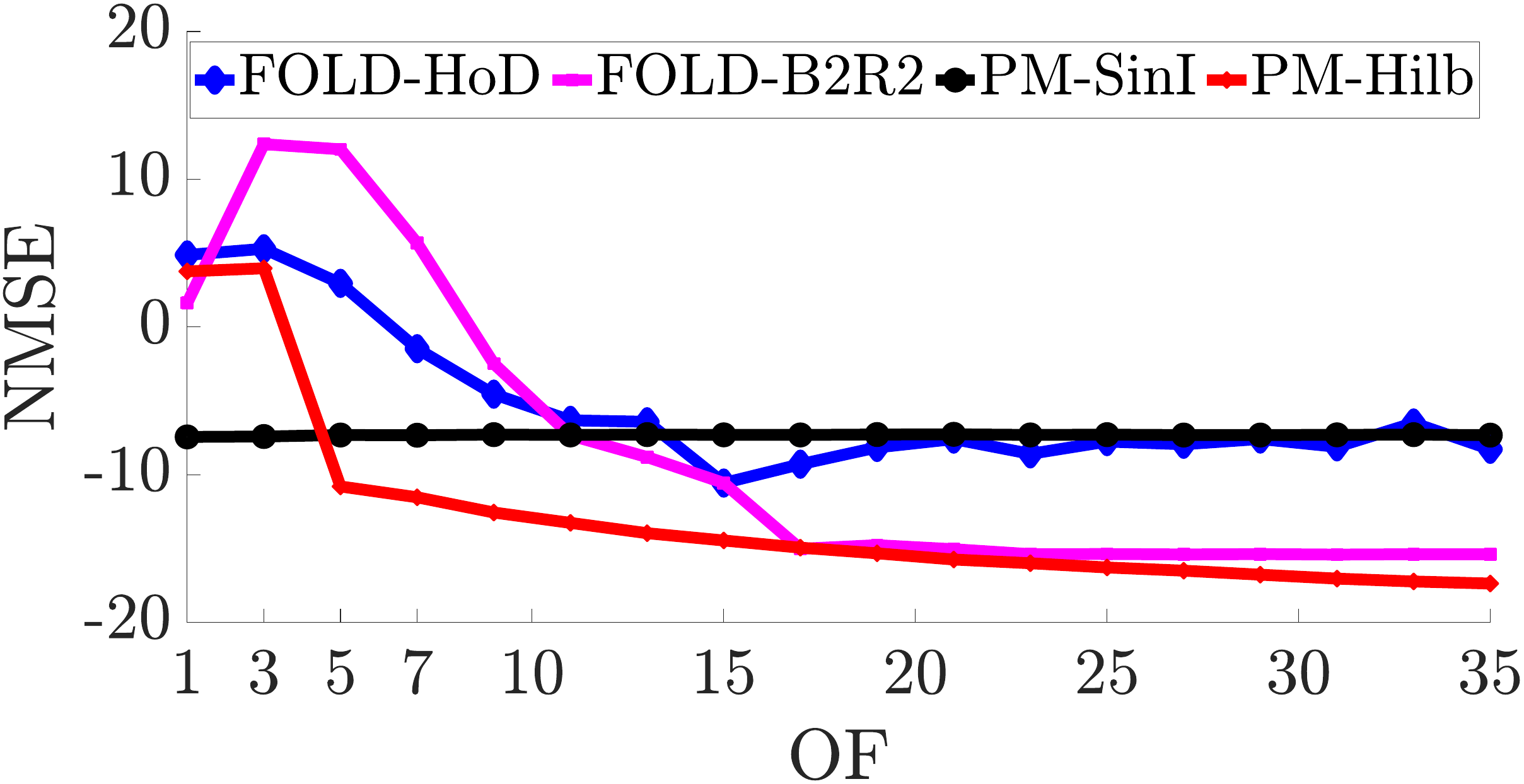}\label{fig:sigma02}}
\subfigure[$\lambda = 0.3$, $\sigma = 0.4\lambda$]{\includegraphics[width=1.7in]{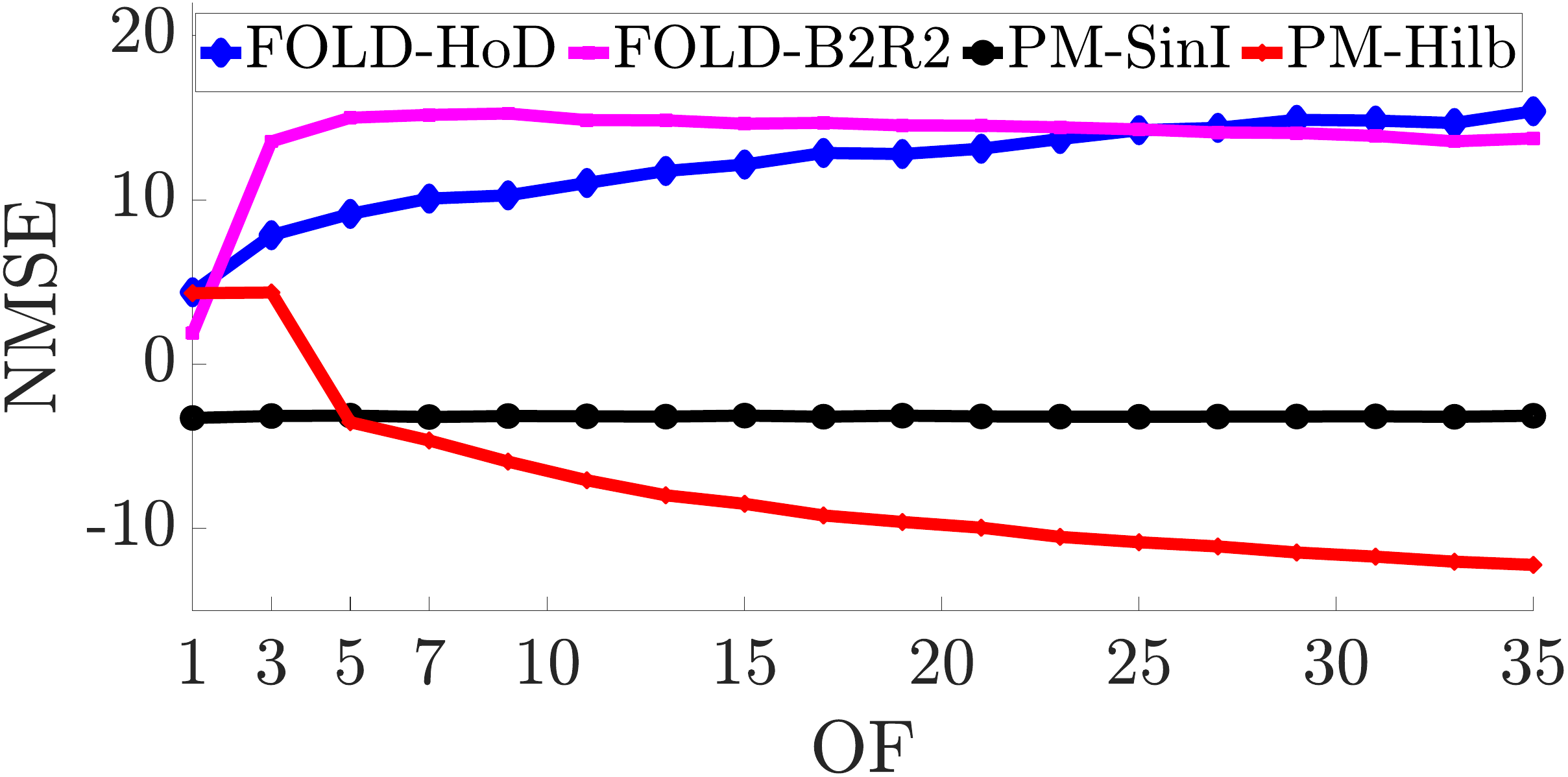}\label{fig:sigma04}}
\end{tabular}
%\end{array}
\caption{Performance comparison of unfolding and DPD algorithms for different $\lambda$s and noise levels. DPD methods can operate at lower OFs compared to unfolding algorithms for a given $\lambda$ and high noise levels.}
\label{fig:errVsOF_diff_sigma}
\end{center}
\end{figure*}

We applied $\sin^{-1}$ and Hilbert-based algorithms for signal estimation from the modulated samples, and for unfolding, we used the higher-order-difference (HoD) \cite{unlimited_sampling17, uls_tsp} and beyond bandwidth residual recovery (B2R2) algorithms \cite{eyar_icassp, eyar_tsp}. The HoD algorithm is one of the first algorithms suggested for unfolding samples of the bandlimited signals, and the B2R2 algorithm is shown to be robust and operates at lower OFs. The signal samples were recovered using these four algorithms, and NMSEs were averaged for 200 independent noise realizations. The errors in the absence of noise for $\lambda = 0.05$ for different OFs are shown in Fig.~\ref{fig:NMSEvsOF_nonoise}. We note that the Hilbert-based DPD algorithm, HoD-based unfolding, and B2R2 algorithm require OFs above 3, 5, and 25 to reconstruct the signal from samples measured by an ADC  with a 20 times lower dynamic range than the signal. However, $\sin^{-1}$-based DPD does not require any OF and operates at the Nyquist rate. This shows the advantage of the proposed modulation-based HDR-ADC.

The low sampling rate requirements of the DPD algorithms compared to unfolding methods were also apparent in the presence of noise, as shown in Fig.~\ref{fig:errVsOF_diff_sigma}. We considered three DRs with $\lambda = 0.05$, $0.1$, and $0.3$,  and four increasing noise levels with $\sigma/\lambda = 0.05, 0.1, 0.2,$ and $0.4$. We found that unfolding algorithms, HoD and B2R2, as noise levels increase, require higher OFs to keep the NMSE acceptable. On the other hand, DPD methods can operate at lower OFs. Moreover, NMSE in the Hilbert-based DPD approach decreases with OF. The performance of the $\sin^{-1}$-based DPD is invariant to the OFs as it is an instantaneous operation. Further, this approach resulted in the lowest error for any noise level for $\text{OF}\leq 3$. We also observed that for $\lambda \geq 0.1$, the B2R2 algorithm resulted in a lower or similar error than the Hilber-based algorithm when $\sigma/\lambda \leq 0.1$. Among the two DPD algorithms, the results imply the following rule of thumb: use $\sin^{-1}$ if OF sampling is expensive but at the cost of higher NMSE; otherwise, apply the Hilber-based DPD.

These four algorithms were employed to recover signal samples, and the NMSEs were averaged across 200 independent noise realizations. In the absence of noise, errors are depicted for $\lambda = 0.05$ across various sampling rates in Fig.~\ref{fig:NMSEvsOF_nonoise}. Notably, the Hilbert-based DPD algorithm, HoD-based unfolding, and B2R2 algorithm necessitates sampling rates above 3, 5, and 25, respectively, to reconstruct signals from samples obtained by an ADC with a dynamic range 20 times lower than the signal. However, the $\sin^{-1}$-based DPD does not require specific sampling rates and functions at the Nyquist rate. This highlights the advantage of the proposed modulation-based HDR-ADC.

The lower sampling rate demands of DPD algorithms compared to unfolding methods were evident even in the presence of noise, as demonstrated in Fig.~\ref{fig:errVsOF_diff_sigma}. We considered three dynamic ranges with $\lambda = 0.05$, $0.1$, and $0.3$, and four escalating noise levels with $\sigma/\lambda = 0.05, 0.1, 0.2,$ and $0.4$. With increasing noise levels, the unfolding algorithms, HoD and B2R2, required higher sampling rates to maintain NMSE at an acceptable level. Conversely, DPD methods could function at lower rates. Furthermore, in the Hilbert-based DPD approach, NMSE decreased with an increase in sampling rate. The $\sin^{-1}$-based DPD remained unaffected by sampling rates as it is an instantaneous operation. Additionally, this approach exhibited the lowest error for any noise level for $\text{OF}\leq 3$. It was also observed that for $\lambda \geq 0.1$, the B2R2 algorithm resulted in lower or comparable errors compared to the Hilbert-based algorithm when $\sigma/\lambda \leq 0.1$. The results suggested a guideline between the two DPD algorithms: use $\sin^{-1}$ when high sampling rates are costly, albeit with higher NMSE; otherwise, opt for the Hilbert-based DPD.
A few key takeaways from the simulation results are summarized as follows.
\begin{itemize}
    \item The PM-based HDR-ADC is efficient in terms of lower reconstruction error than modulo-ADC.
    \item For a given error, PM-based HDR-ADC requires a lower sampling rate.
    \item In the PM-based HDR-ADC, the sampling rate does not increase with a decrease in $\lambda$ for a given error level. In other words, for a fixed $\lambda$, the signal's DR can be increased without increasing the sampling rate. On the other hand, in modulo-ADC, OF has to increase with the signal's DR for a fixed $\lambda$.
    \item The proposed HDR-ADC is more noise-robust than modulo-ADC.
\end{itemize}

The outcomes and comparisons from the simulations focused on bandlimited signals, yet they have the potential to be broadened to encompass diverse signal categories within the PM-based HDR-ADC framework, particularly with the integration of the $\sin^{-1}$-based DPD algorithm. Conversely, extending these findings to modulo-ADC involves adopting one of two approaches.

One approach involves adapting existing unfolding algorithms, primarily tailored for bandlimited or smooth signals, to accommodate specific signal models. The alternative method entails utilizing a bandlimited kernel and subsequently applying established unfolding techniques. However, these approaches might not consistently yield an optimized sampling framework. For instance, extending unfolding techniques designed for smooth signals to FRI signals was demonstrated to necessitate significant oversampling \cite{mulleti2022modulo}. Moreover, establishing theoretical assurances necessitates intricate mathematical considerations \cite{mulleti2022modulo}.

In contrast, utilizing the second approach, i.e., employing a bandlimited kernel, facilitates the utilization of existing unfolding algorithms with their respective optimal sampling frequencies \cite{bhandari2022back}. Nevertheless, this method might not suit many signal models. For instance, custom-made kernels are essential for sampling signals in shift-invariant spaces, which may not inherently be bandlimited.

Considering the benefits offered by the proposed HDR-ADC, we proceed to demonstrate its application in ECG signal sampling, especially in the presence of undesired artifacts.

\begin{figure}
    \centering
    \includegraphics[width = 3.5 in]{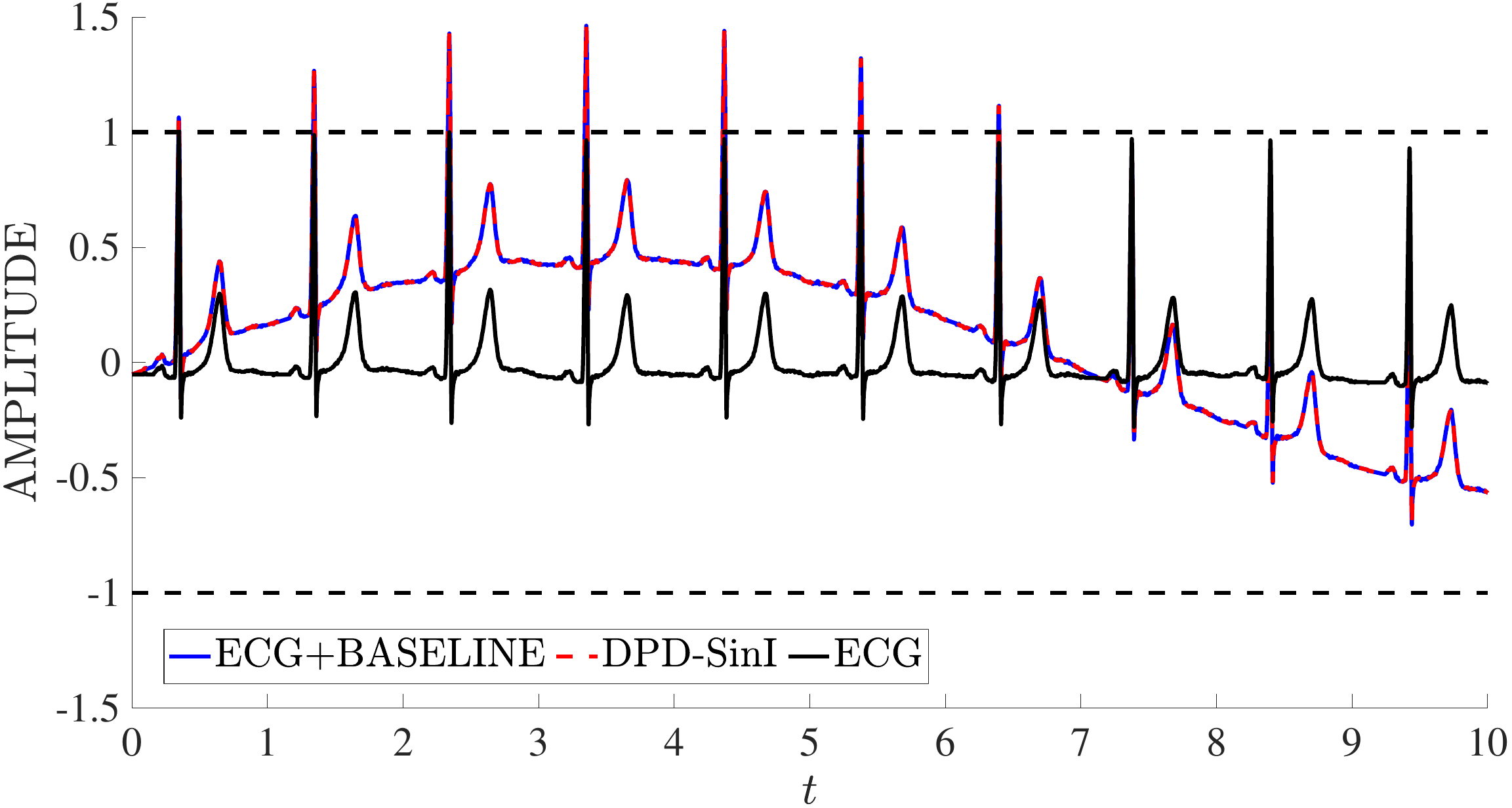}
    \caption{Reconstruction of ECG signal with baseline by using low-DR ADC by using PM.}
    \label{fig:ecg}
\end{figure}

\section{Application to ECG Signals}
\label{sec:ecg}
The simulation outcomes indicated that a modulation-based HDR can function at a reduced sampling rate compared to modulo-HDR. This section highlights the significance of operating at a low sampling rate by exploring a practical application involving the sampling of ECG signals. ECG signals are utilized to gauge various aspects of the heart's condition through a non-invasive process involving the placement of approximately 12 electrodes around the chest. Typically, a standard clinical ECG signal with a bandwidth of under 100 Hz is sampled at or below 500 Hz. Extracting different features from these samples aids in heart condition analysis.

However, a notable challenge during feature extraction lies in the presence of diverse noises contaminating the ECG signals, including power line interference, muscle artifacts, respiration noise, and more. Some of these noises, like muscle artifacts and respiration noise, induce baseline wandering, potentially surpassing the amplitude of the actual ECG signal. Mathematically, if $f(t)$ represents the genuine ECG signal, the signal afflicted by noises can be described as follows:
\begin{align}
\Tilde{f}(t) = f(t) + b(t) + w(t), \label{eq:ECGnoise}
\end{align}
where $b(t)$ accounts for the baseline wandering noise, and $w(t)$ encompasses other noise components. The magnitude of the baseline wandering can significantly exceed that of the ECG signal, that is, $|b(t)|\gg |f(t)|$. To accurately sample $\Tilde{f}(t)$ without encountering clipping, ADC's DR must be sufficiently large to accommodate the unwanted signal $b(t)$. However, utilizing such an HDR-ADC necessitates employing a high number of bits to capture the variations in the ECG signal in the presence of the baseline. 

Alternatively, a combination of modulo-folding and a low-DR ADC could be employed. Yet, oversampling of the signal becomes necessary for unfolding purposes, and the oversampling rate escalates concerning the ratio between the dynamic ranges of $b(t)$ and $f(t)$. This leads to higher bit rates and increased data volume, an undesirable scenario in wearable ECG applications where continuous ECG signal monitoring is essential.

Conversely, employing PM-based DR compression and $\sin^{-1}$-based DPD algorithms does not mandate oversampling. This approach presents an advantageous alternative for managing ECG signal sampling in the presence of baseline wandering and other noise components.

We have showcased the utilization of a low-DR-ADC combined with modulation to capture ECG signals featuring a baseline in Fig.\ref{fig:ecg}. This experimental setup involves an ECG signal obtained from the PhsioNet database, specifically sample number JS00311 from the comprehensive 12-lead electrocardiogram database utilized for arrhythmia studies \cite{physionet1, physionet2}. The signal was sampled at $f_s = 500 $ Hz and initially lacked any baseline artifact. The ECG signal samples are normalized to satisfy $|f(nT_s)|\leq 1$. We introduced a baseline signal given by $b(t) = A_{bl}, \sin(2\pi f_{bl} t),$ where $ f_{bl} = 0.07$ and $A_{bl} = 0.5$. Both the original ECG signal and its version with the added baseline artifact are depicted in Fig.\ref{fig:ecg}. By setting $\lambda = 1$, we applied the proposed PM-based folding and $\sin^{-1}$-based DPD, resulting in exemplary recovery of the ECG signal despite the baseline artifact. Additionally, we attempted modulo folding followed by unfolding algorithms (B2R2 and HoD), assuming the ECG signal was bandlimited to 100 Hz. Unfortunately, both unfolding methods exhibited imperfect results. The HoD algorithm faced challenges due to insufficient oversampling, while the B2R2 algorithm lacked adequate samples within $[-\lambda, \lambda]$. This underscores the advantage of the proposed framework, where augmenting the existing ADC setup to enhance the effective DR is unnecessary. Specifically, the modulation-based dynamic range reduction can be introduced before the current ADC without altering the sampling rate.

Moving on to evaluating the impact of quantization, which holds significance in various applications, we assessed the ECG signal with a baseline possessing a dynamic range of $[-1.5, 1.5]$. Through PM, the signal's DR was reduced to $[-1, 1]$. In both instances, eight bits were utilized for quantization. Employing $\sin^{-1}$-based DPD on the PM samples, we compared the recovered quantized samples with the true ECG samples featuring a baseline. In this scenario, the normalized MSE measured at $-14.7$ dB, while it stood at $-12.2$ dB with a DR of $[-1.5, 1.5]$. This 2.5 dB enhancement highlights the advantage of employing modulation-based HDR over conventional HDR methods.

In summary, modulation-based HDR operates at the Nyquist rate and requires a lower bit count, resulting in a reduced bit rate compared to modulo-ADC or conventional HDR techniques.

\begin{figure}[t]
\centering
\includegraphics[scale=0.5]{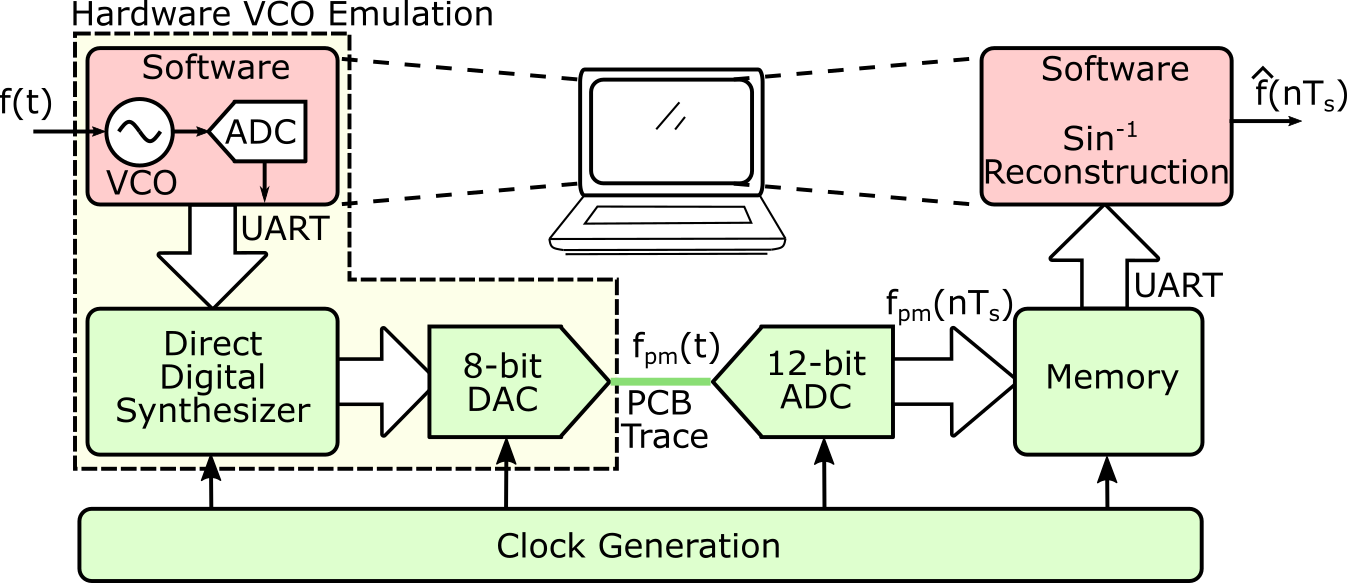}
\caption{Illustration of the hardware for modulation-based high-dynamic range.}
\label{fig:HW_TOP} 
\end{figure}

\begin{figure}[!t]
\centering
\includegraphics[width = 2.5 in]{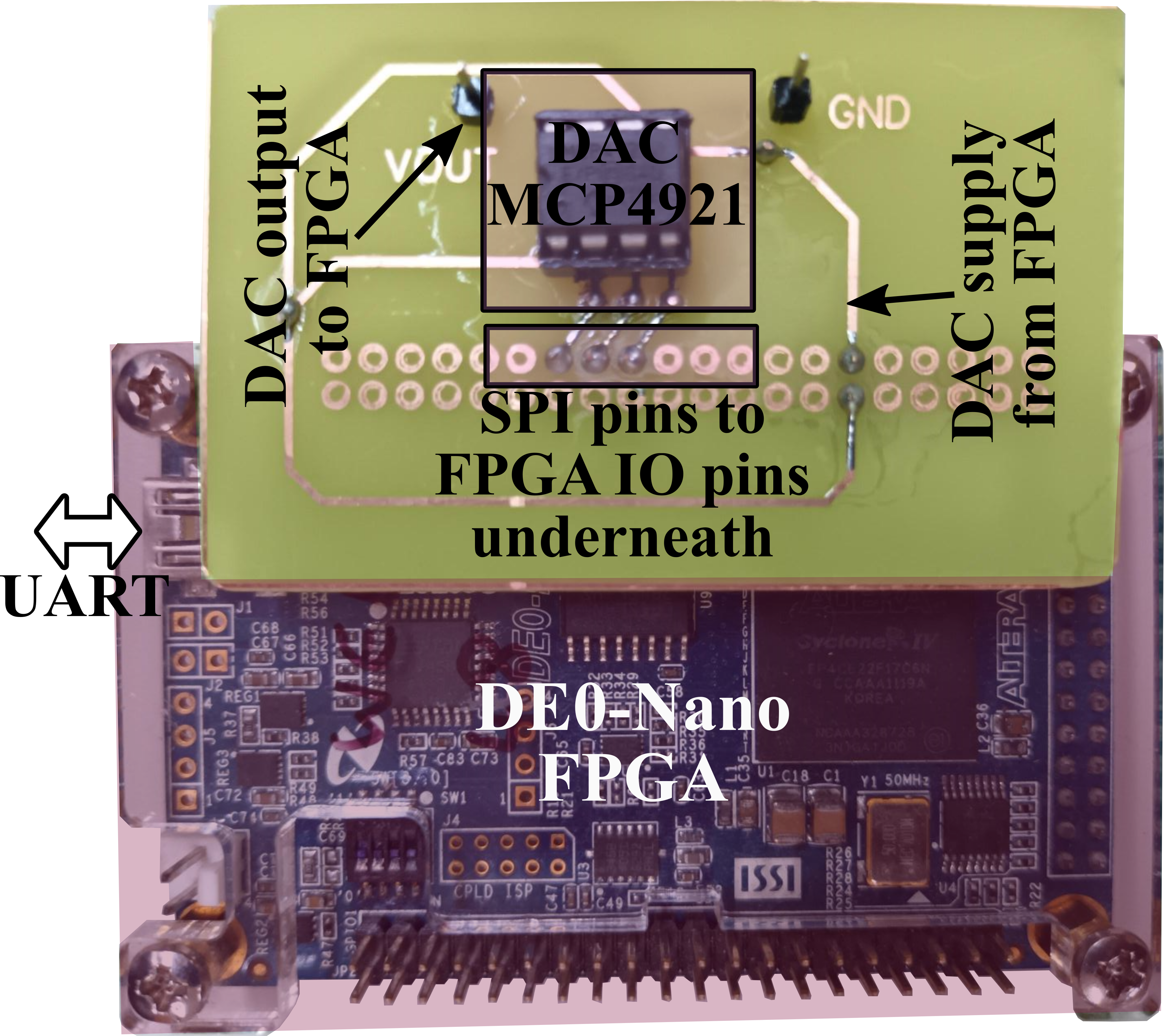}
\caption{Implemented hardware prototype for modulation-based ADC using DE0-Nano FPGA.}
\label{fig:HW} 
\end{figure}

\begin{figure}[t]
\centering
\includegraphics[scale=0.56]{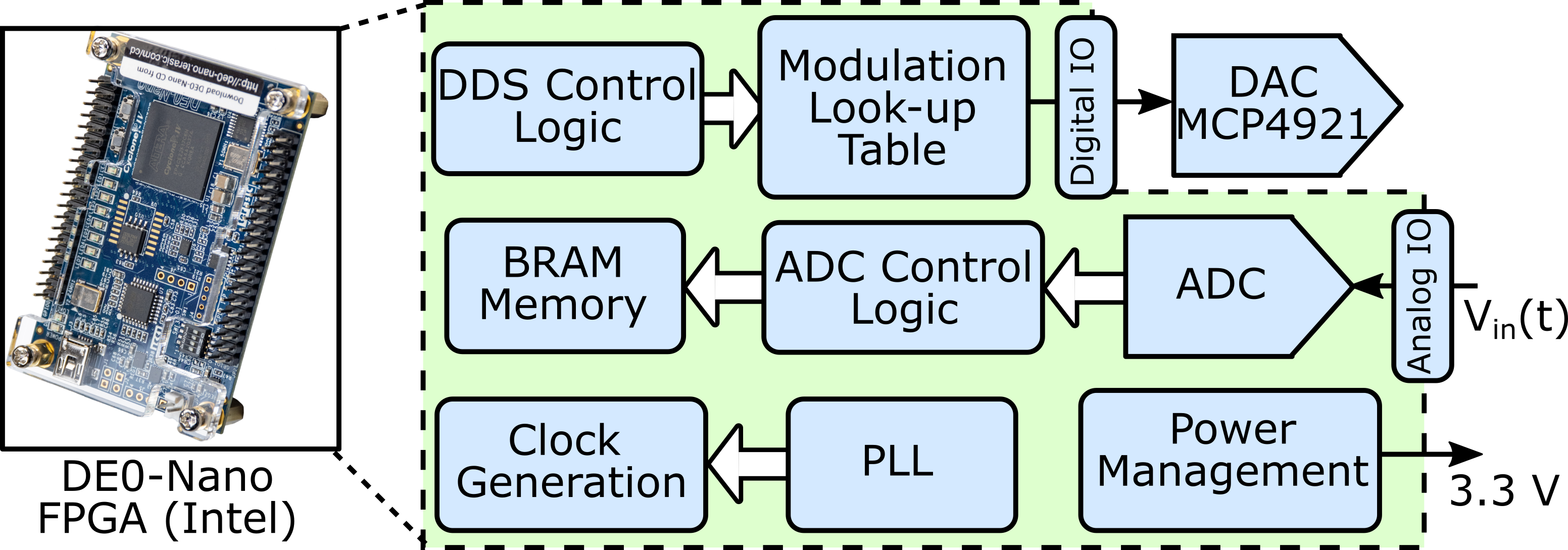}
\caption{Block Diagram of FPGA-based direct-digital synthesizer for modulation and sampling using the internal phase-locked loop (PLL) and ADC.}
\label{fig:HW_BLOCK} 
\end{figure}

\section{Hardware Prototype: Modulation-based HDR}
\label{sec:hw}
We demonstrate the working of the modulation-based HDR using a rapid hardware prototype. In practice, PM is achieved by using a voltage-controlled oscillator (VCO) where a message signal dictates the phase of the oscillations of a sinusoid as in \eqref{eq:fpm}. To enable rapid prototyping, a VCO was emulated in hardware, as discussed next.

\subsection{Modulation-based HDR Hardware Prototype}
The flow diagram of the proposed prototype is shown in Fig.~\ref{fig:HW_TOP}. There are three major components of the prototype: a VCO or a PM signal generator, a low-DR ADC, and a DPD algorithm. The VCO was emulated by using a direct-digital synthesizer (DDS) that was implemented using a field programmable gate array (FPGA). In this work, we have chosen an Intel DE0-Nano board with a Cyclone IV device FPGA~\cite{deo_nano}. The VCO was realized using three components: (a) a software VCO model with quantized output, (b) a direct digital synthesizer logic, and (c) a digital-to-analog converter (DAC). Eight-bit digital samples of PM signal $\fpm(t)$ for a set of sinusoids with varying amplitudes and frequencies were generated through Matlab. This digital data was loaded onto a look-up table (LUT) inside the DDS as illustrated in Fig.~\ref{fig:HW_BLOCK}. The sampled data was converted to the corresponding analog PM signal by an 8-bit DAC. The DAC is followed by an ADC, which samples the PM signal. The ADC and the DAC communicated through a serial-peripheral interface (SPI). The sampled data from the ADC were stored on the FPGA block-RAM (BRAM) and transferred to a computer using universal asynchronous transfer protocol (UART), and the sampled data is reconstructed using $\sin^{-1}$ to obtain $\hat{f}(nT_s)$. The sampling clock for the ADC and the DAC-SPI clocks are generated on the FPGA using onboard phase-locked loops (PLL). 

The hardware prototype was operated with a single-supply voltage source of 3.3 V. The absence of a negative supply voltage means the signal swings were shifted by a common mode voltage of 1.65 V. Within these settings, the DR of the ADC used was set as $[0, 2\lambda]$, where $\lambda = 1.65$ V. This also implies that the DAC used in the VCO had a supply voltage $3.3$ V and the generated PM signal's DR was in the range $[0, 3.3]$ V.

\begin{figure}[!t]
\centering
\includegraphics[width = 3.4 in]{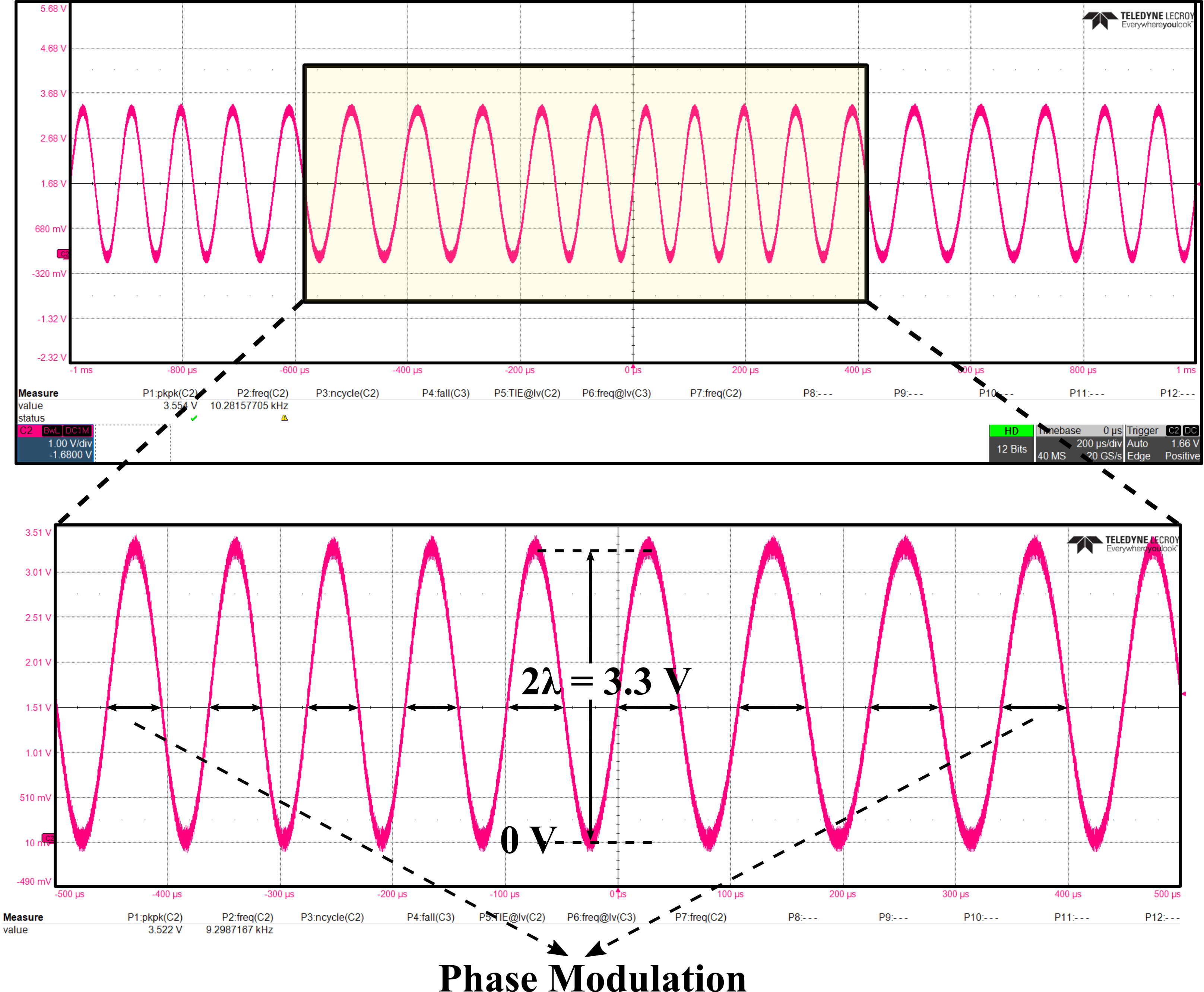}
\caption{Phase modulated waveform from the prototype observed in a digital storage oscilloscope (DSO). Amplitude of the modulated signal = 3.3 V and VCO centre frequency, $f_{c}$ = 10 KHz.}
\label{fig:HW_WAVE}
\end{figure}

The block diagram of the FPGA board that housed the components of the prototype is shown in Fig.~\ref{fig:HW}. The DAC was inserted into the FPGA slot using a custom board. All the required supply and reference voltages for the DAC are obtained from the FPGA board. The prototype communicates with the computer using the UART port shown in Fig.~\ref{fig:HW}.

Next, we compare the sampled and reconstructed signals measured by the hardware prototype. 

\subsection{Experimental Results}
We provide the readers with an output snapshot of the emulated VCO using the DDS-DAC hardware. Fig.~\ref{fig:HW_WAVE} shows the output of the emulated VCO captured on a Lecroy digital storage oscilloscope (DSO). The VCO output is centered around 10 kHz with a phase modulation corresponding with an input signal frequency of $f(t)$ of 70 Hz. To provide a better perspective of the intended phase modulation, a zoomed-in version of the plot is provided in Fig.~\ref{fig:HW_WAVE} clearly showing the DR of 3.3 V and phase modulation in terms of the change in the period of the sinusoidal carrier frequency.\par
\begin{figure*}[!t]
\begin{center}
\begin{tabular}{ccc}
\subfigure[$c = 33$]{\includegraphics[width=2in]{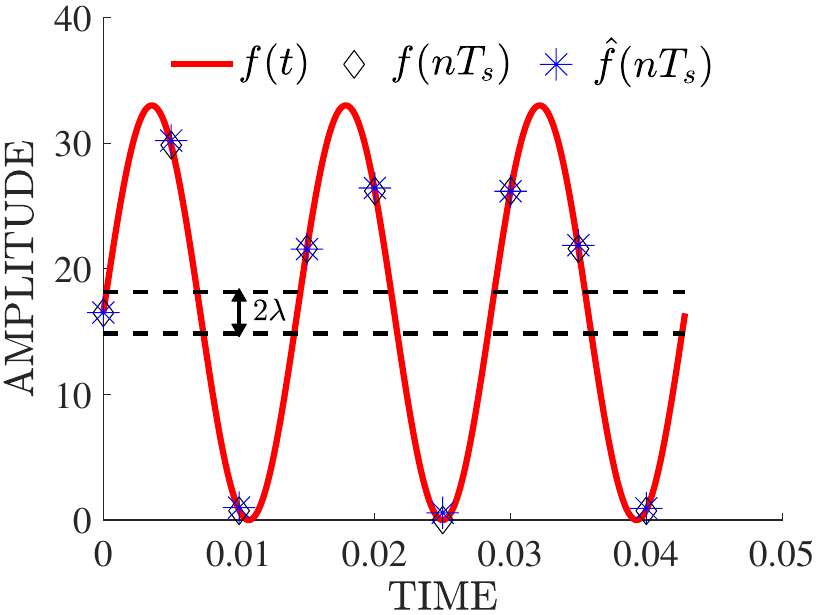}\label{fig:A4}} 
\subfigure[$c=16.5$]{\includegraphics[width=2in]{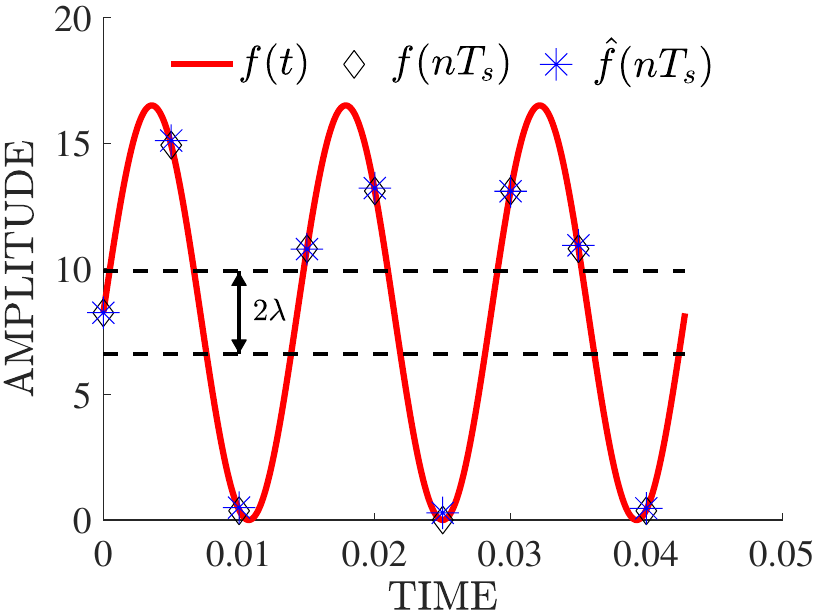}\label{fig:A5}} 
\subfigure[$c=9.9$]{\includegraphics[width=2in]{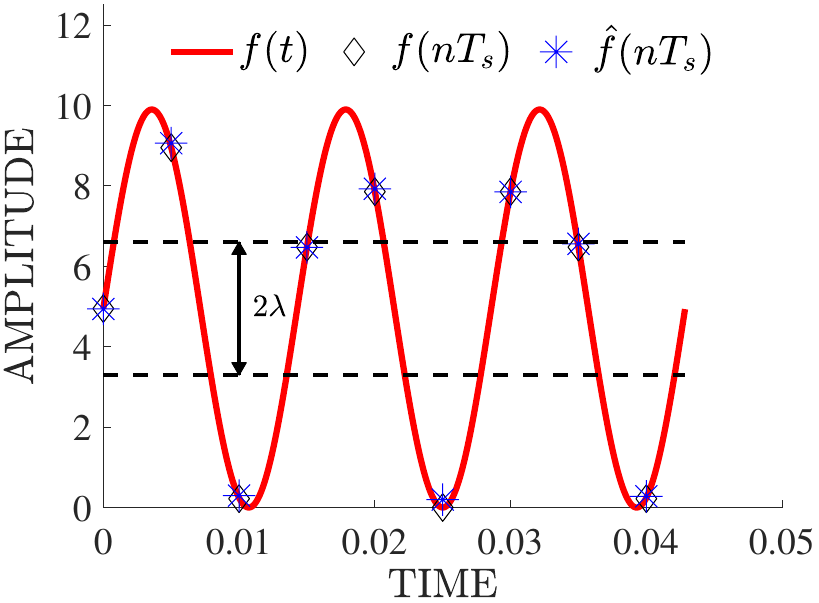}\label{fig:A6}}
\end{tabular}
%\end{array}
\caption{Phase modulation-HDR based sample reconstruction with ADC DR of 2$\lambda$ = 3.3, ADC's sampling rate is 200 Hz, VCO center frequency, $f_{c}$ = 10 KHz, with sinusoidal input $f(t)$ of 70 Hz with dynamic range (a) $c/\lambda$ = 10 (b) $c/\lambda$ = 5 (c) $c/\lambda$ = 3. The true samples and the estimated samples are given as $f(nT_s)$ and $\hat{f}(nT_s)$, respectively.}
\label{fig:HW_RECON1}
\end{center}
\end{figure*}
\begin{figure*}[!t]
\begin{center}
\begin{tabular}{ccc}
\subfigure[$c = 33$]{\includegraphics[width=2in]{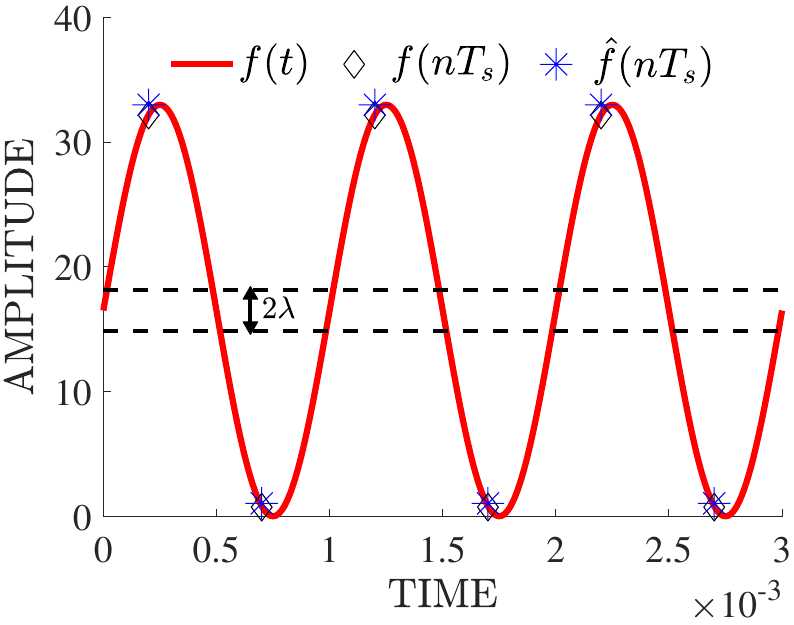}\label{fig:A1}} 
\subfigure[$c=16.5$]{\includegraphics[width=2in]{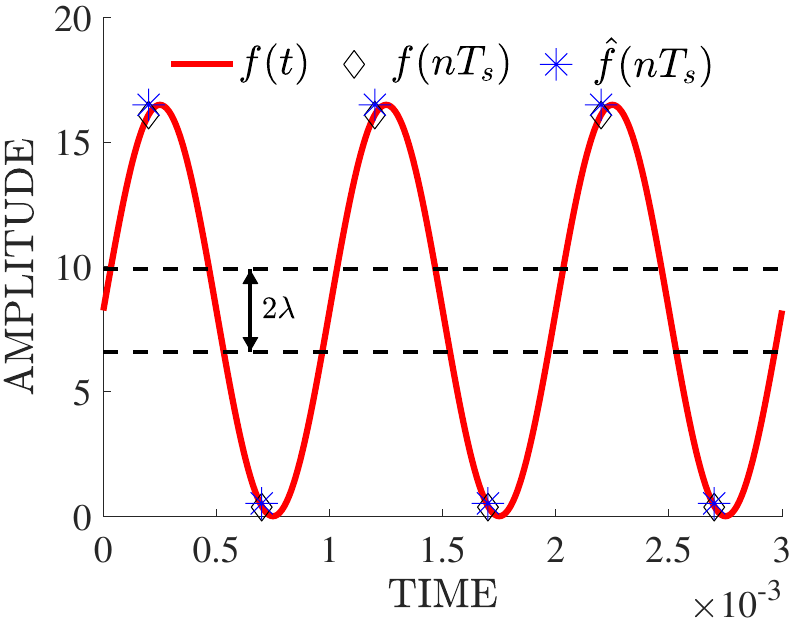}\label{fig:A2}} 
\subfigure[$c=9.9$]{\includegraphics[width=2in]{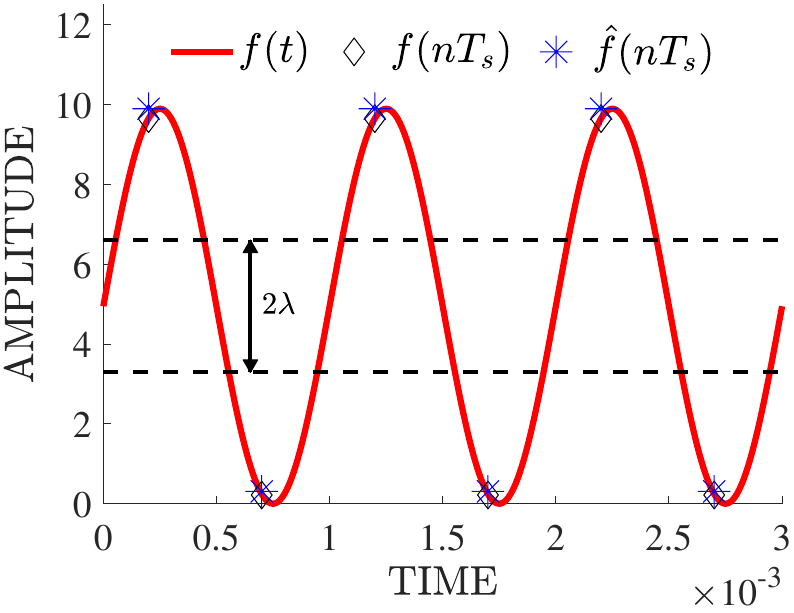}\label{fig:A3}}
\end{tabular}
%\end{array}
\caption{Phase modulation-HDR based sample reconstruction with ADC DR of 2$\lambda$ = 3.3, ADC's sampling rate is 2 kHz, VCO center frequency, $f_{c}$ = 10 KHz, with sinusoidal input $f(t)$ of 1 kHz with dynamic range (a) $c/\lambda$ = 10 (b) $c/\lambda$ = 5 (c) $c/\lambda$ = 3. The true samples and the estimated samples are given as $f(nT_s)$ and $\hat{f}(nT_s)$, respectively.}
\label{fig:HW_RECON2}
\end{center}
\end{figure*}

Next, we provide reconstruction results of the data captured through the prototype for two cases: (a) sinusoidal input signal $f(t)$ with 70 Hz frequency with three different $c/\lambda$ of 10, 5, and 3 with ADC sampling frequency of 200 Hz (near Nyquist) (b) sinusoidal input signal with 1 kHz frequency with three different dynamic range ratio, $c/\lambda$ of 10, 5 and 3 with ADC sampling frequency of 2 kHz (at Nyquist). For both cases, the VCO center frequency is chosen as 10 kHz. The reconstruction results, along with the ideal instant samples, are shown in Fig.~\ref{fig:HW_RECON1} and Fig.~\ref{fig:HW_RECON2}. The sinusoidal signal $f(t)$ is shown in red. The true samples $f(nT_s)$ and the estimated samples $\hat{f}(nT_s)$ are shown using diamond-shaped pointers and blue stars, respectively. We noted near-perfect reconstruction of the at or above Nyquist rates of the signals from which the original signal can be reconstructed. 

In the results presented, the signal's DR is chosen to be ten times larger than the ADC's DR. Since the emulated VCO hardware in the presented prototype has a software sampling with DDS, as shown in Fig.~\ref{fig:HW_TOP}, practically, there is no limit on the DR of the input signal. While realizing the hardware with an analog VCO circuit, a few other parameters of the VCO need to be considered to decide the maximum DR of the input signal. Similarly, the input signal's bandwidth could be as large as possible, provided that the ADC can sample it at the Nyquist rate.

It must be noted that a modulo-ADC realized by using a feedback loop, as in \cite{mulleti2023hardware}, requires a DAC or its equivalent in the feedback loop. The task of the feedback DAC is to generate a piecewise constant signal $z(t) \in 2\lambda \mathbb{Z}$ such that the folded signal, 
\begin{align}
    f_\lambda(t)=f(t)+z(t),
    \label{eq:flambda}
\end{align}
lies within the dynamic range $[-\lambda, \lambda]$. The maximum value of $z(t)$ is given as $2(k+1)\lambda$ where $k$ is an integer satisfying the inequality $(2k+1)\lambda < c \leq (2k+3) \lambda$. More precisely, we have that $|z(t)| \leq |c-\lambda|$. Hence, for a fixed $\lambda$, the output voltage of the DAC and, hence, DAC's power supply should be as large as $\pm(c-\lambda)$. At the same time, the ADC in the modulo framework requires a power supply of $\pm \lambda$. As a result, a modulo-ADC requires multiple and large power sources to sample HDR signals. Further, modulo-ADC requires oversampling to recover the true samples $f(nT_s)$ from the folded samples $f_\lambda(nT_s)$. Oversampling leads to higher power consumption \cite{walden_adc}. 

In comparison, the proposed PM-based HDR-ADC requires a single power source with voltages $\pm \lambda$ or $2\lambda$. Further, additional power due to oversampling is not required by operating at the Nyquist rate. Hence, the proposed framework requires a lower power budget than a modulo-ADC.

In the proposed hardware prototype, due to emulated VCO, there is no limit on the dynamic range of input signal that is required to be modulated. However, when the proposed hardware is implemented using an off-the-shelf VCO, which is ongoing work, 
 to keep the signals' DR as large as possible, additional circuitry is required to ensure that the $\mu$ is below the threshold values. 

\section{Conclusions}
In this paper, we proposed a PM-based HDR-ADC that operates at the Nyquist rate for bandlimited signals or at the minimum rate for any class of signals. Theoretical bounds on the identifiability results and practical algorithms were discussed. Simulation results and application to ECG signals underscore the proposed framework's significance compared to the modulo-ADC. Further, the hardware prototype of the proposed HDR-ADC was discussed, which shows Nyquist rate sampling and reconstruction of signals, which are ten times larger than the ADC's DR. 

\bibliographystyle{IEEEtran}
\bibliography{US_biblios,refs,refs2,TheoryPart/refs_hw}

\end{document}